\documentclass[a4paper,11pt]{article}
\pdfoutput=1
\usepackage{amsmath}
\usepackage{amssymb}
\usepackage{cite}
\usepackage{tabularx}
\usepackage{graphicx}
\usepackage{cancel}
\usepackage{feynmp}
\usepackage{xspace}
\usepackage{flafter}
\usepackage{ifpdf}

\def\spa#1.#2{\left\langle#1\,#2\right\rangle}
\def\spb#1.#2{\left[#1\,#2\right]}
\def\spaa#1.#2.#3{\langle\mskip-1mu{#1}
                  | #2 | {#3}\mskip-1mu\rangle}
\def\spbb#1.#2.#3{[\mskip-1mu{#1}
                  | #2 | {#3}\mskip-1mu]}
\def\spab#1.#2.#3{\langle\mskip-1mu{#1}
                  | #2 | {#3}\mskip-1mu\rangle}
\def\spba#1.#2.#3{\langle\mskip-1mu{#1}^+
                  | #2 | {#3}^+\mskip-1mu\rangle}
\def\spav#1.#2.#3{\|\mskip-1mu{#1}
                  | #2 | {#3}\mskip-1mu\|^2}
\def\jc#1.#2.#3{j^{#1}_{#2#3}}
\def\blfootnote{\xdef\@thefnmark{}\@footnotetext} 
\newcommand{\Ca}{\ensuremath{C_{\!A}}\xspace}

\newcommand{\nf}{\ensuremath{n_{\!f}}\xspace}
\newcommand{\Nc}{\ensuremath{N_{\!C}}\xspace}
\newcommand{\gs}{\ensuremath{g}}

\newcommand{\HEJ}{{\tt HEJ}\xspace}
\newcommand{\epsfig}[1]{} 
\title{\begin{normalsize}
\begin{flushright}
Edinburgh 2012/13\\
IPPP/12/45, DCPT/12/90\\
CP3-Origins-2012-017, DIAS-2012-18
\end{flushright}
\end{normalsize}
\vspace*{1cm}
W Plus Multiple Jets at the LHC with High Energy Jets}
\author{Jeppe~R.~Andersen$^{a}$, Tuomas Hapola$^{b}$,
  Jennifer~M.~Smillie$^{c}$\\\mbox{}\\
  $^a$ Institute for Particle Physics Phenomenology,\\University of Durham, Durham, DH1 3LE, UK\\
  $^b$ {CP}$^{ \bf 3}${-Origins} \& DIAS, Campusvej 55, DK-5230 Odense M, Denmark.\\
  $^c$ School of Physics and Astronomy, University of Edinburgh,\\
         Mayfield Road, Edinburgh EH9 3JZ, UK.}
\date{June 28, 2012}
\begin{document}
\maketitle
\begin{abstract}
  We study the production of a $W$ boson in association with $n$ hard QCD
  jets (for $n\ge 2$), with a particular emphasis on results relevant for the
  Large Hadron Collider (7~TeV and 8~TeV). We present predictions for this
  process from High Energy Jets, a framework for all-order resummation of the
  dominant contributions from wide-angle QCD emissions. We first compare
  predictions against recent ATLAS data and then shift focus to observables
  and regions of phase space where effects beyond NLO are expected to be
  large.
\end{abstract}

\newpage
\tableofcontents

\section{Introduction}
\label{sec:introduction}

There is already a wealth of analyses of data from the Large Hadron Collider (LHC),
many of which are beginning to stress-test the perturbative descriptions of proton-proton collisions at LHC energies. It is of
course essential to demonstrate to what level the Standard Model (SM) processes
are understood. This is often
the first step towards a discovery (or limit) on new physics. 

A particularly interesting process is $W$ plus jets where, in the leptonic decay
mode, the neutrino leads to events with missing transverse energy.  The
combination of multiple
jets and missing
energy is a very common signal in models beyond the Standard Model, arising
from a production mechanism with QCD charged particles followed by decay
chains ending with a (semi-)stable, weakly interacting particle.

There have been a number of data analyses of this channel already by both the
ATLAS~\cite{Aad:2010pg,Aad:2012en} and CMS~\cite{Chatrchyan:2011ne} experiments.
The production of a $W$ boson in association with jets is also a clean
environment for QCD studies, and as such could be an important testing ground
for experimental methods used to study methods like e.g.~jet vetos. What is
learned in $W$ plus jets could be applied in searches for the Higgs boson and
other new physics.

There has also been a great deal of recent theoretical progress in
fixed-order calculations of this
channel,
where the state-of-the-art is now $W$ plus four jets in the leading colour
approximation~\cite{Berger:2010zx}, or $W$ plus three jets in the full colour
dressing~~\cite{KeithEllis:2009bu,Berger:2009zg,Berger:2009ep}. The NLO calculations lead to very good
agreement for total cross sections and, for the right scale choices, also for
several differential distributions.

Much work has been reported on estimating effects beyond NLO. The NLO
calculations for $W$-production in association with up to
two~\cite{Frederix:2011ig} and three jets~\cite{Hoeche:2012ft} have been
systematically merged with a parton shower. Furthermore,
LoopSim~\cite{Rubin:2010xp} has also been applied to approximate higher
fixed-order contributions to $W$ production in regions where the $K$-factor
is large.

In this study we present the application of the \emph{High Energy Jets}
(\HEJ) formalism developed in Refs.~\cite{Andersen:2009nu,Andersen:2009he} to
the process of $W$-production (and leptonic decay) in association with
\emph{at least} two hard jets. The general formalism has previously been
applied to pure multi-jet production~\cite{Andersen:2011hs}, with the results
from the accompanying flexible Monte Carlo implementation used in several
analyses of LHC
data\cite{Aad:2011jz,Collaboration:2012gw,Chatrchyan:2012pb}. The all-order
results of \HEJ are complementary to those obtained from a parton shower
approach. The simplifications made by \HEJ to the perturbative series to
allow all-order results to be obtained become exact in the limit of large
invariant mass between all particles. This corresponds to the limit where
also the (B)FKL amplitudes\cite{Fadin:1975cb,Kuraev:1976ge,Kuraev:1977fs} are
exact. The production of $W$+dijets was previously studied within the BFKL
formalism\cite{Andersen:2001ja}. Compared to this, the present formalism
introduces several major improvements: it relaxes several kinematic
assumptions made in the earlier study, improves the accuracy of the
resummation, and introduces matching to fixed order results.

In pure \HEJ, there are no collinear singularities, no shower and
no hadronisation: the output is a pure partonic calculation. However, this can all be consistently included by merging the
results with a shower Monte Carlo by a procedure which avoids
double-counting\cite{Andersen:2011zd}. Such double-counting would otherwise
arise in particular in the soft regions, which are treated in both \HEJ and a
parton shower.

This paper is structured as follows.  In Section~\ref{sec:anatomy-w-jets} we
study the various production channels for $W$ plus jets at the LHC and
properties of the jet radiation.  In Section~\ref{sec:w-bosons-hej} we
summarise the construction of the HEJ amplitudes and their implementation in
a fully flexible Monte Carlo. The resulting program is available at
\textsc{http://cern.ch/hej}.  In Section~\ref{sec:comparison-lhc-data} we
compare predictions for kinematic distributions to LHC data, before embarking
in Section~\ref{sec:results} on a discussion of regions of phase space and
observables where effects beyond NLO are expected to be large. We end in
Section~\ref{sec:conclusions} with a brief discussion.


\boldmath
\section{The Anatomy of $W$ Plus Jets at the LHC}
\label{sec:anatomy-w-jets}
\unboldmath

In this section, we focus on the production  of a $W^{+/-}$-boson (followed
by a leptonic decay) in association with at least two jets. We will from now
denote both $W^+$ and $W^-$ as $W$, and both electron and positron as
$e$. Furthermore, we will apply the following acceptance cuts:
\begin{align}
  \begin{split}
    \label{eq:anatcuts}
    p_{\perp j}>30\ {\rm GeV}, \quad &\left| \eta_j \right|<4.4, \quad
    p_{\perp e}>20\ {\rm GeV}, \quad \left|\eta_{e}\right| <
    2.5, \quad
    \cancel{E}_T>25\ {\rm GeV},\\
    M_{\perp W}&= \sqrt{2\ p_{\perp e}\ p_{\perp \nu}\
      (1-\cos(\phi_e-\phi_\nu))}\ > 40\ {\rm GeV},
  \end{split}
\end{align}
where $\eta$ is the pseudorapidity and $\phi$ is the azimuthal angle of the
respective particle momentum.

The kinematic requirements of the two jets and the charged lepton alone
ensure that the parton density functions are probed only for a light-cone
momentum fraction $x$ larger than $3.4\cdot 10^{-4}$ at 7~TeV ($1.7\cdot
10^{-4}$ at 14~TeV). The neutrino momentum will contribute further to the
momentum fraction, and thus the dynamics is well within the regime where
standard, collinearly factorised pdfs and hard scattering matrix elements
describe the cross sections accurately. We will therefore concentrate on
perturbative higher order corrections within this framework, and not discuss
small-$x$ issues like $k_\perp$-factorisation or unintegrated pdfs. We will
use the standard MSTW2008NLO\cite{Martin:2009iq} pdf set throughout. The
accompanying program is interfaced to \texttt{LHAPDF}, so any publicly
available pdf set can be used.


In earlier studies\cite{Andersen:2003gs,Andersen:2011hs,Andersen:2011zd} we
established a strong connection between the average number of hard jets in
events, and the rapidity difference between the most forward/backward hard
jet in pure jet production. This is a result not just of the opening of phase
space for the emission of additional jets, but also the possibility of a
colour octet exchange between particles ordered in rapidity. Such a correlation was recently
confirmed by data\cite{Aad:2011jz}.

It is obviously relevant to discuss the extent to which the behaviour
observed in pure jets is relevant for the production of $W$+dijets. The
explanation from
BFKL\cite{Fadin:1975cb,Kuraev:1976ge,Kuraev:1977fs,Balitsky:1978ic} for the
correlation between the average number of jets and the rapidity span of the
events\cite{Andersen:2003gs} relies on the higher order corrections to the
dijet processes which allow for a colour octet exchange between all
rapidity-ordered partons. 

\begin{figure}[btp]
  \centering
  \includegraphics[width=0.5\textwidth]{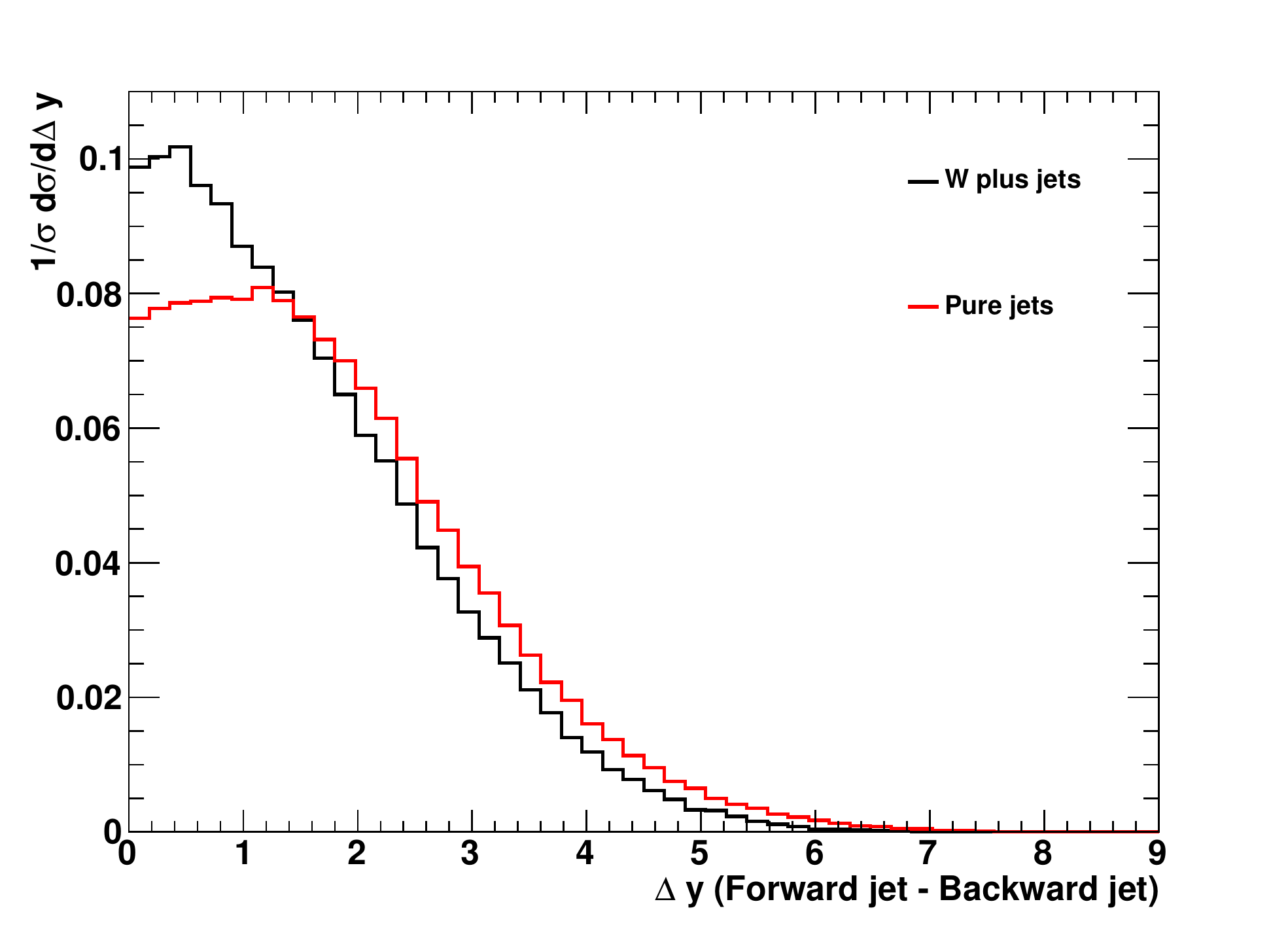}
  \caption{The normalised distribution of the rapidity difference between the
    two jets, $\mathrm{d}\sigma/\mathrm{d}(y_f-y_b)$, for $W$ plus two jets at
    tree-level (black), within the cuts of Eq.~\eqref{eq:anatcuts}.  Also shown
    for comparison is the same distribution for dijet production, with
    equivalent cuts on the jets.}
  \label{fig:Wqrapdif}
\end{figure}

\begin{table}[btp]
\begin{center}
  \begin{tabular}{|l||c|}
\hline
Subprocess & $\sigma_{2j}\ [pb]$ \\
\hline
$q q' \to W q q' $ & 11.6 \\
$qg \to W qg  $ & 67.7 \\
$q \bar{q} \to W q \bar{q} $ & 9.9  \\
\hline
$q\bar{q} \to W gg,  W q'\bar{q}'$ & 5.8  \\
$gg \to W q\bar{q}  $ & 4.3 \\ 
\hline
\end{tabular}
\end{center}
\caption{Leading order cross sections (in pb) for the production of a $W$ boson in
  association with two jets for the LHC at $\sqrt{s}=7$~TeV. The subprocesses above
  the horizontal line are those which allow a colour octet exchange
  between the two initial state (and the two final state) partons, while those below do not.}
\label{table:Wrapidity}
\end{table}
In Table~\ref{table:Wrapidity} we have listed the sub-process contributions to
the cross section for $Wjj$-production at leading order within the cuts of
Eq.~\eqref{eq:anatcuts}. The processes above the horizontal line are those which
allow colour octet exchange between the two initial (and the two final state) partons. Clearly, these dominate the
cross section.  
Such colour exchanges will dominate the higher order corrections in the
limit of large rapidity separation between all partons. However, this
requirement also means that quarks are produced only as the partons which are
extremal in rapidity in the scattering. One could worry that the cut on the
centrality of the charged lepton from the $W$ decay would force the quark-line
emitting the $W$ to also be central, thereby suppressing the phase space for a
large rapidity span between the jets.  In figure~\ref{fig:Wqrapdif} we plot
$1/\sigma \mathrm{d}\sigma/\mathrm{d}(y_f-y_b)$ for $W+$dijets at leading order.
Also shown for comparison is the equivalent distribution for pure dijets.  While
the $W+$dijets distribution is slightly more peaked at zero, there is very
little difference in the shape of the distributions above a rapidity
span of around 1.5,
indicating that a systematic resummation of radiation between jets is just as
relevant for $W$+dijets as for dijets.

The analyses presented above demonstrate that at the LHC, the sub-processes
allowing for colour octet exchange in the $t$-channel dominate the rate for
$W$+dijets.  This observation will be the guiding principle for the resummation for
$W$+jets implemented in \HEJ.


\boldmath
\section{$W$ Plus Jets in High Energy Jets}
\label{sec:w-bosons-hej}
\unboldmath

The framework of \emph{High Energy Jets}
(\HEJ)\cite{Andersen:2009nu,Andersen:2009he} constructs explicit
approximations to the real and virtual corrections to the perturbative hard
scattering matrix element at any order. Furthermore, matching to the full
tree-level results for the first few higher order corrections are implemented
through a jet merging algorithm. The framework, and the application to the
description of $W$+jets is described in this section.

\subsection{All-Order Amplitudes}
\label{sec:jet-production-high}

The starting point for the \HEJ approach is an approximation to the
hard-scattering matrix element for $2\to n$ partons plus the leptonic 
decay products of the  $W$. This is built from the
dominant terms in the High Energy (or Multi-Regge Kinematic) limit, which is
defined as:
\begin{align}
  \label{eq:HElimit}
  \begin{split}
    &\forall i,j\quad s_{ij}\to \infty \quad |p_{i\perp}| \sim |p_{j\perp}| \\
    {\rm or}\ {\rm equivalently}\quad & i \in\{1,n-1\} \quad y_{i-1}\ll y_i\ll y_{i+1} \quad
    |p_{i\perp}| \sim |p_{j\perp}|,
  \end{split}
\end{align}
where $i,j$ label the $n$ final state quarks and gluons, ordered in rapidity.
There is no constraint on the momenta of the $e,\nu$.
\begin{figure}[btp]
  \centering
  \begin{minipage}[b]{0.25\textwidth}
    \includegraphics[width=\textwidth]{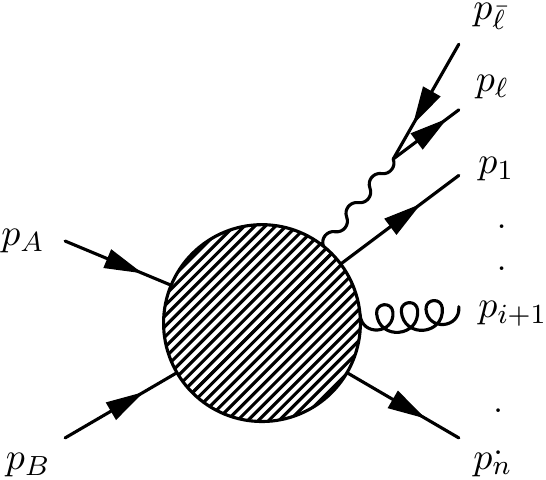}
    \vspace{1cm}
  \end{minipage} 
  \begin{minipage}[b]{0.1\textwidth}
    \centering
    $\longrightarrow$

    \vspace{3cm}
  \end{minipage}
  \includegraphics[width=0.6\textwidth]{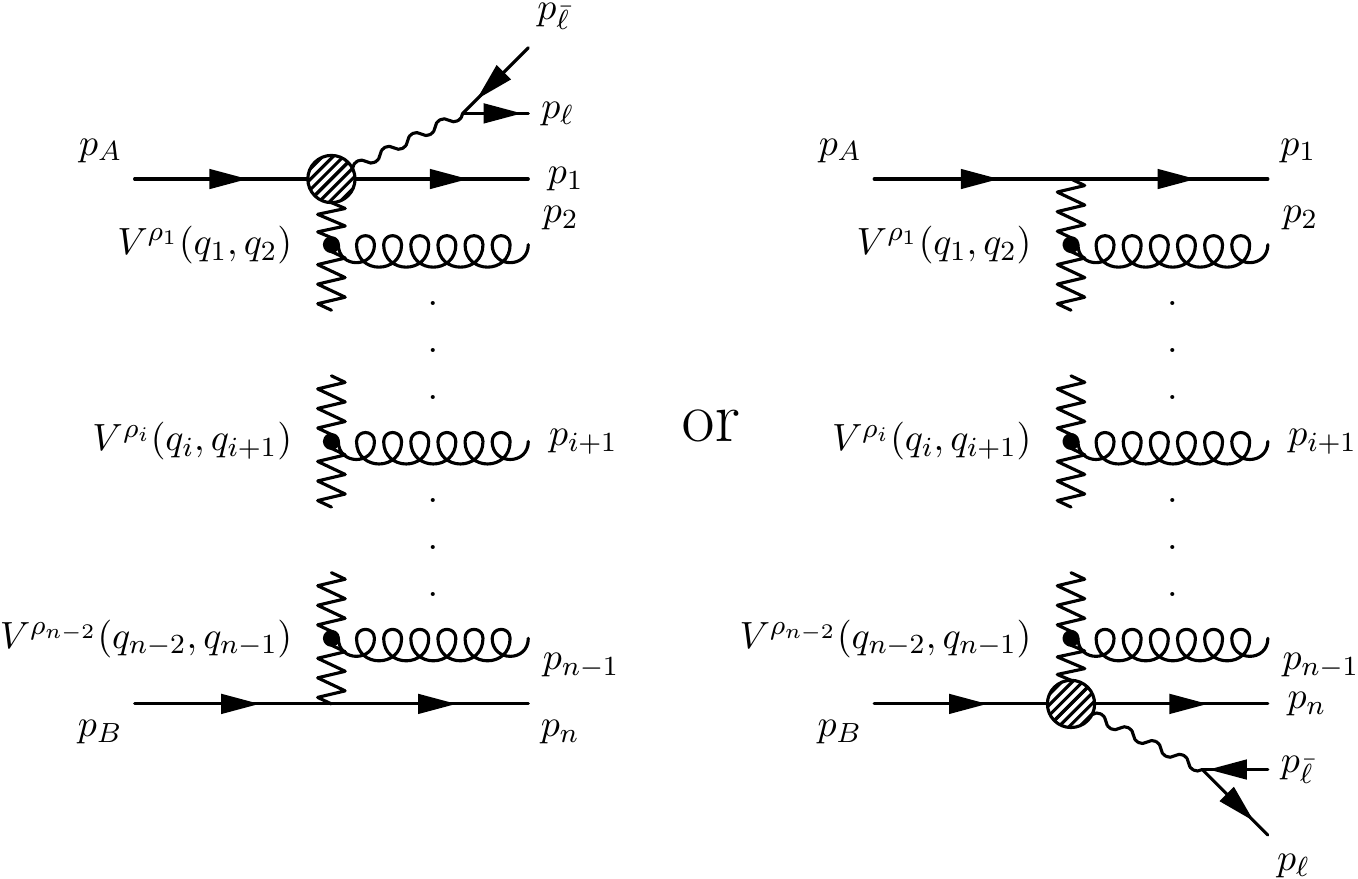}
  \caption{Sample kinematics for $qQ\to q^\prime g \ldots gQ$.  The diagrams on
    the right-hand side show the dominant kinematic configurations in the High
    Energy limit where the outgoing partons are drawn and numbered in order of
    increasing rapidity.  The effective vertices, $V(q_i,q_{i+1})$, are defined
    in Eq.~\eqref{eq:Veff}.}
  \label{fig:genericWjets}
\end{figure}
In this limit, the scattering amplitude is dominated by the poles in the
$t$-channel momenta, as depicted in Fig.~\ref{fig:genericWjets} and defined as
\begin{align}
  \label{eq:qidef}
  q_i=p_A-(p_e-p_\nu)\cdot \delta-p_1-\ldots-p_{i-1},
\end{align}
where $\delta=1$ if the $W$ is interpreted as emitted from the quark line
with momenta $p_A,p_1$, and $\delta=0$ otherwise. The numbering of momenta is according to decreasing rapidity, $A$ indicates
the forward moving incoming parton, $B$ the backward moving one. The leading
contributions to the poles arise from the particle and momentum
configurations which allow for a colour octet exchange between each
neighbouring (in rapidity) pair of partons. We will call these \emph{FKL}
configurations. For example, in dijet production $ug\to ug$ and $ud\to uggd$
are FKL configurations with that rapidity order, while $ug\to gu$ and $ud\to
us\bar sd$ are not.  Each relaxation of a colour ordering induces a relative
suppression of $1/s$ in the squared matrix element and hence is subdominant
in the High Energy limit. The all-order treatment in \HEJ currently only
describes FKL configurations. Other contributions are included order by order
with standard tree-level matrix elements.

Scattering amplitudes factorise according to rapidity in the High Energy
limit. Based on the work of
Ref.\cite{Andersen:2009nu,Andersen:2009he,Andersen:2011hs} we choose a form
of amplitudes for each helicity configuration, which is a) gauge invariant,
b) exact in the MRK limit, and c) sufficiently fast to evaluate that a
numerical integration over the phase space of the sum over any number of
emissions can be accurately evaluated. For the case at hand, the tree-level
approximation to the square of the matrix element is
\begin{align}
  \label{eq:MHEJtree}
  \begin{split}
    \overline{\left|\mathcal{M}_{\rm HEJ}^t(\{ p_i, p_e,p_\nu\})\right|}^2 = \ &\frac 1 {4\
       (\Nc^2-1)}\ \left\|S_{f_a f_b\to f_1 f_n\ p_e\ p_\nu}\right\|^2\\
     &\cdot\ \left(g^2\ K_{f_1}\ \frac 1 {t_1}\right) \cdot\ \left(g^2\ K_{f_n}\ \frac 1
       {t_{n-1}}\right)\\
     & \cdot \prod_{i=1}^{n-2} \left( {g^2 C_A}\
       \left(\frac {-1}{t_it_{i+1}} V^\mu(q_i,q_{i+1})V_\mu(q_i,q_{i+1})\right)\right)\\
  \end{split}
\end{align}
The bold-face $\mathbf{p}_i$ indicates the transverse components, with $\mathbf{p}_i^2>0$.
Each part of this equation will now be explained in full
detail. $\left\|S_{f_a f_b\to f_1 f_n\ p_e\ p_\nu}\right\|^2$ is the helicity
sum of the contracted currents
\begin{align}
  \label{eq:combinecouplings}
    ||S_{ud\to d \nu_\ell \bar \ell d}||^2 &= \frac{g_W^4}{4} \left| \frac 1 {
      p_W^2-M_W^2+i\ \Gamma_W M_W}\right|^2\sum_{h_b,h_2} |S_{ud\to d
      \nu_\ell \bar \ell d}^{- h_b \to - h_2 - -}|^2,
\end{align}
where we have picked the case of the extremal partons to represent the
scattering $ud\to d \nu_\ell \bar \ell d$ as an example, and the incoming
$u$-quark as the forward-moving parton (i.e.~with momentum $p_a$). The spinor
contraction for each helicity is simply
\begin{align}
  \label{eq:Wspinorstring}
  S^{-h_b\to -h_2--}_{ud\to dd\nu_\ell \bar \ell}=j_{W\mu}(i,\ell,\bar \ell,
  o)\  g^{\mu\nu}\ \bar u^{h_2}(p_2)\gamma_\nu u^{h_b}(p_b)
\end{align}
with
\begin{align}
  \label{eq:Weffcur}
  &j_W^\mu(i,\ell,\bar \ell, o) = \bar u^-(p_{\rm out}) \left( \gamma^\alpha
    \frac{\cancel{p}_W+\cancel{p}_{\rm out}}{(p_W+p_{\rm out})^2}\ \gamma^\mu + \gamma^\mu
    \frac{\cancel{p}_{\rm in}-\cancel{p}_W}{(p_{\rm in}-p_W)^2}\ \gamma^\alpha \right)
    u^-(p_{\rm in})\ \cdot\ \bar{u}^-(p_\ell) \gamma_\alpha
    u^-(p_{\bar\ell}).\nonumber\\
   & \hspace{1cm} \begin{minipage}[b]{0.2\textwidth}
     \includegraphics[width=\textwidth]{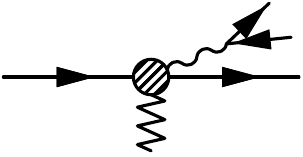}
   \end{minipage}
   \begin{minipage}[b]{0.1\textwidth}
     \centering{=}
    \vspace{0.7cm}
   \end{minipage}
    \begin{minipage}[b]{0.2\textwidth}
      \includegraphics[width=\textwidth]{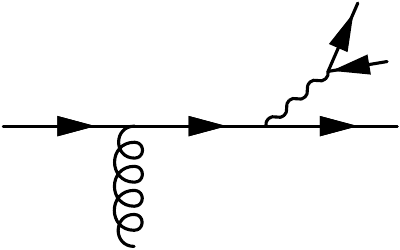}
    \end{minipage}
\begin{minipage}[b]{0.1\textwidth}
    \centering{+}
    \vspace{0.7cm}
  \end{minipage}
    \begin{minipage}[b]{0.2\textwidth}
  \includegraphics[width=\textwidth]{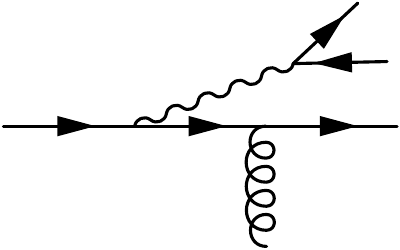}    
\end{minipage}
\end{align}
This is simply the spinor representation of a fermion line emitting a
(decaying) W and a (possibly off-shell) gluon, as represented by the figure
in the equation above.

In $qg$-initiated processes, the $W$ is obviously attached to the quark
line. However, for $qQ$-initiated processes, the $W$ emission can be assigned
to one or both lines. However, the formalism developed needs the $W$-emission
to be assigned to just one line. Therefore, we 
assign the $W$ to either quark line $a$ or $b$ on an event-by-event basis according 
to the ratio:
\begin{align}
  \label{eq:Wemitratio}
  a:b \quad = \qquad e^{-|y_a-y_W|}:e^{-|y_b-y_W|}, 
\end{align}
so that the quark closest in rapidity to the $W$ is favoured, but not chosen
exclusively. The event weight is adjusted so the outcome is independent of
this choice. Picking one specific quark line allows for the definite assignment of $t$-channel
momenta in Eq.~\eqref{eq:qidef} and the calculation of the squared amplitude
in Eq.~\eqref{eq:combinecouplings}\footnote{This approach does not include the (kinematically
suppressed) effects of
quantum interference from the emission of $W$ in the cases where there are
two incoming quark lines of same flavour)}.

For a quark line, the colour factor $K_{f_i}$ is just $C_F$. We will use the
results of Ref.\cite{Andersen:2009he} to include the case of $qg\to q'g\cdots gg
\ell\bar\ell$ by simply using a colour factor depending on the light-cone
momenta fraction $z=p_n^-/p_b^-$ of the scattered gluon:
\begin{align}
  \label{eq:Kg}
  K_g=\left[ \frac 1 2\ 
       \frac{1+z^2}{z}\ \left(C_A -\frac 1 {C_A}\right)+
       \frac{1}{C_A}\right].
\end{align}
It can be seen that $K_g\to C_A$ for $p_n^-/p_b^-\to 1$, in agreement with the well-known result
in this limit\cite{Combridge:1984jn}.

The $t$-channel propagator denominators are given by $t_i=q_i^2$, with $q_i$
defined in Eq.~\eqref{eq:qidef}. Additional (gluon) emissions are approximated
with a string of effective emission vertices~\cite{Andersen:2009nu}, given by:
\begin{align}
  \label{eq:Veff}
    \begin{split}
  V^\rho(q_i,q_{i+1})=&-(q_i+q_{i+1})^\rho \\
  &+ \frac{p_A^\rho}{2} \left( \frac{q_i^2}{p_{i+1}\cdot p_A} +
    \frac{p_{i+1}\cdot p_B}{p_A\cdot p_B} + \frac{p_{i+1}\cdot p_n}{p_A\cdot p_n}\right) +
  p_A \rightarrow p_1 \\ 
  &- \frac{p_B^\rho}{2} \left( \frac{q_{i+1}^2}{p_{i+1} \cdot p_B} + \frac{p_{i+1}\cdot
      p_A}{p_B\cdot p_A} + \frac{p_{i+1}\cdot p_1}{p_B\cdot p_1} \right) - p_B
  \rightarrow p_n
  \end{split}
\end{align}
where $p_g=q_{i+1}-q_i$ is the momentum of the emitted gluon.
This simple structure makes it very fast to integrate in an efficient
phase-space generator~\cite{Andersen:2008ue,Andersen:2008gc}.  This allows the
number of final state particles to be treated as a variable in the integration.
The four-momenta of all final state particles is available in every event,
allowing arbitrary cuts and analyses.

An infrared pole will be generated from the phase space integration
over the region $|p_{g,i}|\to 0$ for each $i$. In these regions,  
\begin{align}
  \label{eq:subtraction}
  \frac {-1}{t_i t_{i+1}}V^\mu(q_i,q_{i+1})V_\mu(q_i,q_{i+1}) \to \frac{4}{\mathbf{p}_i^2}.
\end{align}
The simple $4/\mathbf{p}_i^2$-term is therefore used as a real-emission
subtraction term (with the integral form added to the virtual corrections). Infrared poles are also generated by the inclusion of virtual corrections to
the $t$-channel gluon propagator factors. These are accounted for by the Lipatov
Ansatz\cite{Balitsky:1978ic}, which replaces the standard gluon propagator
factor $1/t_i$ with
\begin{align}
  \label{eq:LipatovAnsatz} \frac 1 {t_i}\ \to\ \frac 1 {t_i}\ \exp\left[\hat
\alpha (q_i)(y_{i-1}-y_i)\right]
\end{align} with
\begin{align} 
  \hat{\alpha}(q_i)&=-\gs^2\ \Ca\
  \frac{\Gamma(1-\varepsilon)}{(4\pi)^{2+\varepsilon}}\frac 2
  \varepsilon\left({\bf q}^2/\mu^2\right)^\varepsilon.\label{eq:ahatdimreg}
\end{align} 
We organise the cancellation of the poles from these two sources using
dimensional regularisation, the subtraction term confined to a small region
of real emission phase space with ${\bf p}_i^2<\lambda^2$. Further details in
Ref.\cite{Andersen:2011hs}.

The final expression for the all-order regularised square of the scattering
matrix element is then given as
\begin{align}
  \label{eq:MHEJ}
  \begin{split}
    \overline{\left|\mathcal{M}_{\rm HEJ}^\mathrm{reg}(\{ p_i, p_e,p_\nu\})\right|}^2 = \ &\frac 1 {4\
       (\Nc^2-1)}\ \left\|S_{f_a f_b\to f_1 f_n\ p_e\ p_\nu}\right\|^2\\
     &\cdot\ \left(g^2\ K_{f_1}\ \frac 1 {t_1}\right) \cdot\ \left(g^2\ K_{f_n}\ \frac 1
       {t_{n-1}}\right)\\
     & \cdot \prod_{i=1}^{n-2} \left( {g^2 C_A}\
       \left(\frac {-1}{t_it_{i+1}} V^\mu(q_i,q_{i+1})V_\mu(q_i,q_{i+1}) -
         \frac{4}{\mathbf{p}_i^2}\ \theta\left(\mathbf{p}_i^2<\lambda^ 2\right)\right)\right)\\
     & \cdot \prod_{j=1}^{n-1} \exp\left[\omega^0(q_j,\lambda)(y_{j-1}-y_j)\right], \\
     \omega^0(q_j,\lambda)=\ &-\frac{\alpha_s N_C}{\pi} \log\frac{{\bf q}_j^2}{\lambda^2}.
  \end{split}
\end{align}

\subsection{Merging with Fixed Order}
\label{sec:merging}

The regularised matrix elements in Eq.~\eqref{eq:MHEJ} allow for an
integration over the momenta of all particles. The cross section is then
obtained as
\begin{align}
  \begin{split}
    \label{eq:resumdijet}
    \sigma_{W+2j}^\mathrm{resum}=&\sum_{f_1, f_2}\ \sum_{n=2}^\infty\
    \prod_{i=1}^n\left(\int_{p_{i\perp}=0}^{p_{i\perp}=\infty}
      \frac{\mathrm{d}^2\mathbf{p}_{i\perp}}{(2\pi)^3}\ 
      \int \frac{\mathrm{d} y_i}{2}
    \right)\
    \int \frac{\mathrm{d}^3p_e}{(2\pi)^3\ 2 E_e}\ 
    \
    \int \frac{\mathrm{d}^3p_\nu}{(2\pi)^3\ 2 E_\nu}\ 
    \\
    &\frac{\overline{|\mathcal{M}_{\mathrm{HEJ}}^{\mathrm{reg}}(\{ p_i,p_e,p_\nu\})|}^2}{\hat s^2} \cdot \ x_a f_{A,f_1}(x_a, Q_a)\
    \cdot x_2
    f_{B,f_2}(x_b, Q_b)\\
    &\times (2\pi)^4\ \delta^2\!\!\left(\sum_{k=1}^n
      \mathbf{p}_{k\perp}+p_{e\perp}+p_{\nu\perp}\right )\ \mathcal{O}_{2j}(\{p_i\}),
  \end{split}
\end{align}
where the first sum is over the flavours $f_1, f_2$ of incoming partons.  In
order to stay within the relevant kinematics for the formalism, we require
the two extremal partons to be part of the extremal hard jets identified by
the jet clustering algorithm. Furthermore, we require at least two such hard
jets to be present. These requirements are implemented by the jet observable
$\mathcal{O}_{2j}$ in Eq.~\eqref{eq:resumdijet}. The distribution of any
observable can be obtained by simply binning the cross section in
Eq.~\eqref{eq:resumdijet} in the appropriate variable formed from the
explicit momenta. Obviously, multi-jet rates can also be calculated by
multiplying by further multi-jet observables $\mathcal{O}_{3j},
\mathcal{O}_{4j},\ldots$.

The simple structure of the \HEJ framework makes it feasible to merge with
other theoretical descriptions, where relevant.  By default, \HEJ contains
matching to fixed-order tree-level amplitudes in two different ways.
Firstly, for flavour and momentum configurations arising in the resummation
described above, the approximation to the $n$-jet production can be
reweighted to full tree-level accuracy. This is achieved by a merging
procedure which clusters all the
$m$ momenta generated by the resummation into $n$ on-shell momenta
$\left\{p^\mathrm{new}_{\mathcal{J}_l}(\{p_i\})\right\}$ close to the
reconstructed jet momenta\cite{Andersen:2011hs}. The event weight is then adjusted with the ratio
of the full $n$-jet matrix element (evaluated by
\texttt{MadGraph}~\cite{Alwall:2007st}) to the approximate one obtained as
the $\alpha_s^n$-expansion of Eq.~\eqref{eq:MHEJ} to the $n$-jet production
rate (obtained from Eq.~\eqref{eq:MHEJtree}). Currently,
reweighting up to $n=4$-jets is applied. The reweighted resummed cross section is then
found as
\begin{align}
  \begin{split}
    \label{eq:resumdijetmerged}
    \sigma_{W+2j}^\mathrm{resum,merged}=&\sum_{f_1, f_2}\ \sum_{n=2}^\infty\
    \prod_{i=1}^n\left(\int_{p_{i\perp}=0}^{p_{i\perp}=\infty}
      \frac{\mathrm{d}^2\mathbf{p}_{i\perp}}{(2\pi)^3}\ 
      \int \frac{\mathrm{d} y_i}{2}
    \right)\
    \int \frac{\mathrm{d}^3p_e}{(2\pi)^3\ 2 E_e}\ 
    \
    \int \frac{\mathrm{d}^3p_\nu}{(2\pi)^3\ 2 E_\nu}\ 
    \\
    &\frac{\overline{|\mathcal{M}_{\mathrm{HEJ}}^{\mathrm{reg}}(\{ p_i,p_e,p_\nu\})|}^2}{\hat s^2} \cdot \ x_a f_{A,f_1}(x_a, Q_a)\
    \cdot x_2
    f_{B,f_2}(x_b, Q_b)\\
    &\times\ \sum_{m=2}^4 \mathcal{O}_{mj}^e(\{p_i\})\ w_{m-\mathrm{jet}}\\
    &\times (2\pi)^4\ \delta^2\!\!\left(\sum_{k=1}^n
      \mathbf{p}_{k\perp}+p_{e\perp}+p_{\nu\perp}\right )\ \mathcal{O}_{2j}(\{p_i\}),
  \end{split}
\end{align}
with $\mathcal{O}_{mj}^e$ the exclusive $m$-jet observable and 
\begin{align}
  \label{eq:matchfact}
  w_{n-\mathrm{jet}}\equiv\frac{\overline{\left|\mathcal{M}^{\mathrm{Tree}}\left(\left\{p^\mathrm{new}_{\mathcal{J}_l}(\{p_i\})\right\}\right)\right|}^2}{\overline{\left|\mathcal{M}_\mathrm{HEJ}^{t}\left(\left\{p^\mathrm{new}_{\mathcal{J}_l}(\{p_i\})\right\}\right)\right|}^2}.
\end{align}

The event sample arising from the equation above is finally supplemented with
the kinematic configurations (up to $n=4$-jets) not arising in this
resummation. This is obviously a very na\"ive matching, and improvements
along a CKKW-L\cite{Catani:2001cc,Lonnblad:1992tz} procedure should be
pursued. Alternatively, the \HEJ resummation could be expanded to cover 
configurations formally subleading in the MRK limit.

\begin{figure}[btp]
  \centering
  \includegraphics[width=0.5\textwidth]{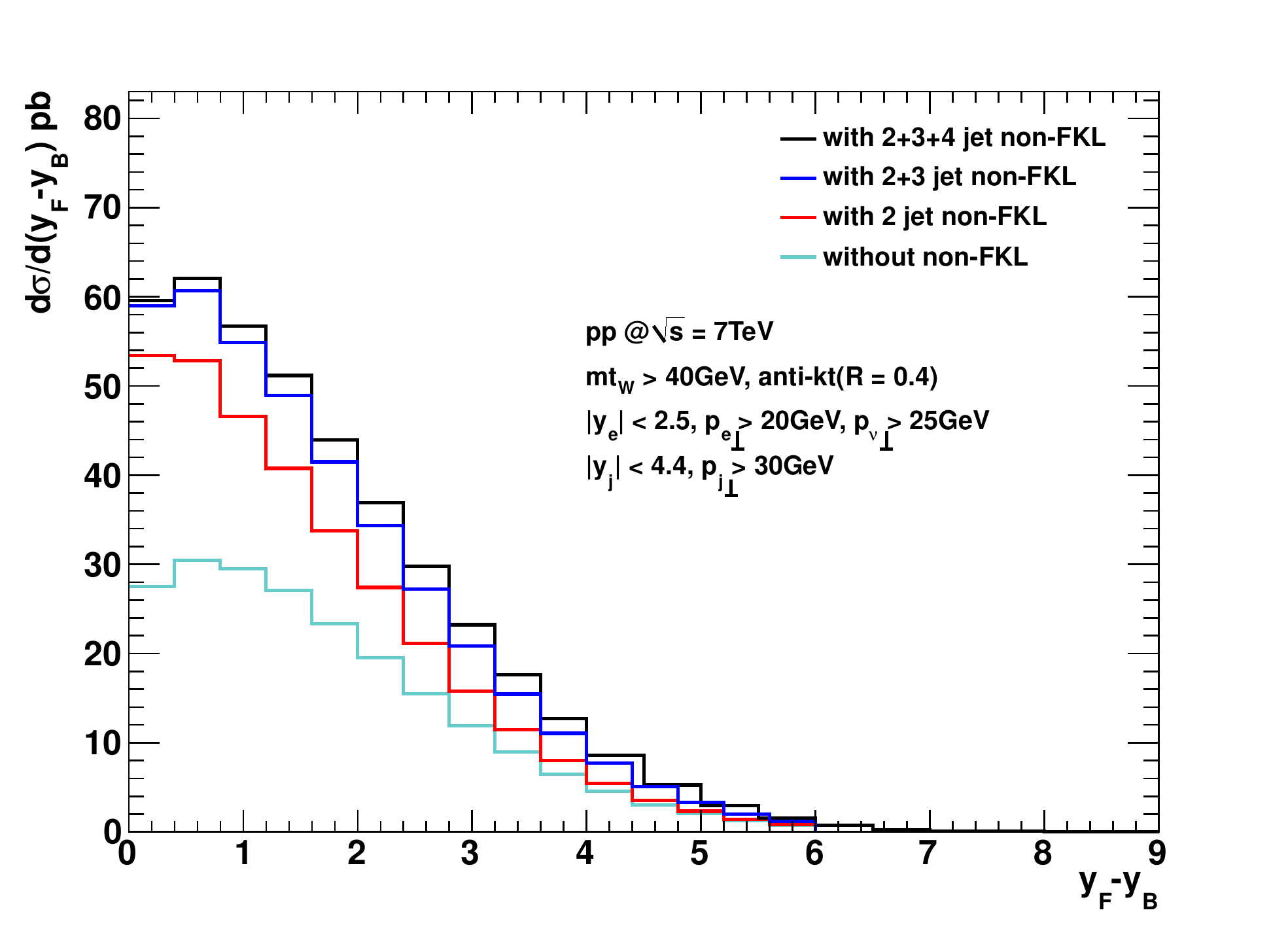}
  \caption{The rapidity span distribution from \HEJ for inclusive $W$ plus dijet
    production for the FKL configurations (turquoise), with 2-jet non-FKL added
    (red), then with 3-jet non-FKL matching added (blue) and lastly also
    including 4-jet non-FKL merging.}
  \label{fig:matching}
\end{figure}

In the meantime, we can assess the importance of the fixed-order contributions
with the present implementation.  Figure~\ref{fig:matching} shows the rapidity
span distribution including different levels of matching.  Clearly at small
rapidity spans, the non-FKL 2-jet matching is dominating the cross-section,
demonstrating the importance of including these contributions.  The impact
diminishes as the rapidity span increases (as expected).  The further addition
of non-FKL 3- and 4-jet matching (blue and black lines) is visible but less
significant for this variable.

Another important direction for the development of \HEJ is work
on merging with parton showers.  \HEJ is constructed to approximate the emission
of QCD particles at wide angles.  While this includes emissions with transverse
momentum as low as 1~GeV, it does not include a description of collinear
emissions.  This can be added by consistently merging with a parton shower
program.  However, this must be done with some care to avoid double-counting
soft emissions.  A subtraction scheme to merge with the \textsc{Ariadne} parton
shower~\cite{Lonnblad:1992tz} has been developed~\cite{Andersen:2011zd}, and
further work is ongoing to merge with other parton shower implementations.

The resummation and merging procedures discussed so far do not rely on a
particular choice for the factorisation and renormalisation scales.  We have
implemented four different scale choices for $\mu_R=\mu_F$ (although they in
principle do not need to be set equal) as follows. We list them here with the
numbering scheme applied in the input file for the program:\newpage
\begin{itemize}
\item[0:] Fixed scale of your choice, set in ``scale''.
\item[1:] $H_T/2$, where $H_T$ is the transverse sum of all final state particles
  including $p_{\perp \ell^\pm}$ and $\cancel{E}_T=p_{\perp \nu}$.\footnote{Note
    that this is strictly greater than the value that can be measured in the
    detector, where instead the momenta of the jets is added (not the original
    partons).}
\item[2:] $p_{\perp {\rm max}}$ -- the maximum $p_\perp$ of any single jet in each event.
\item[3:] The geometric mean of the identified hard jets, $\left( \prod_{j=1}^n p_{\perp j} \right)^{1/n}$.
\end{itemize}

In addition, there is the option to add logarithmic corrections, mimicking
the part of the NLL BFKL corrections which are proportional to the LL
kernel. These can be included by setting the
``logcorrect'' parameter in the input file to 1. These corrections modify $\omega_0(q,\lambda)$ of
Eq.~\eqref{eq:MHEJ} to give instead
\begin{align}
  \begin{split}
    \label{eq:NLLtrajectoryrun}
    \omega_0(q_j,\lambda) =&\ - \frac{\alpha_s(\mu^2)\
      \Ca}{\pi}\ln\left(\frac{\mathbf{q}_j^2}{\lambda^2}\right)\ \left(
      1+\frac{\alpha_s(\mu^2)}2 \frac{\beta_0}{4 \pi}\ln\frac{\mu^4}{\mathbf{q}_j^2\lambda^2}
    \right),\\
    \beta_0=&\frac{11}3\Nc - \frac 2 3 \nf
  \end{split}
\end{align}
while for the real emission vertices the coupling is multiplied by
\begin{align}
  \label{eq:NLLrealemissionreg}
  \left ( 1 - \alpha_s(\mu^2) \frac{\beta_0}{4\pi}\ln \mathbf{p}_i^2/\mu^2\right).
\end{align}
See Ref.~\cite{Andersen:2011hs} for a full discussion.  For the results
presented in the next two sections we have used $\mu_R=\mu_F=H_T/2$
and included the logarithmic corrections associated with the scale. This
last choice is motivated by reducing the impact of the NLL corrections.
These choices are not an indication of an optimised fit to data, and 
other scale choices could be studied.


\section{Comparison to LHC data}
\label{sec:comparison-lhc-data}

Having outlined the description of $W$ plus jets in \HEJ, in this section we
compare the resulting predictions to the data we have from the LHC.  A recent
ATLAS study of $W$ production in association with jets~\cite{Aad:2012en} gave
results for many interesting distributions, using the full 2010 data sample
of 36~pb$^{-1}$.  The cuts used in that study, and therefore here, match
those of section~\ref{sec:anatomy-w-jets} (eq.~\eqref{eq:anatcuts}), with the
addition of an isolation cut, $\Delta R(\ell, j) > 0.5$, applied to all jets.
Throughout, the \HEJ predictions are shown together with a band indicating
the variation found when varying the renormalisation and factorisation scale
by a factor of two in either direction. Obviously, this variation is only a
rough indication of the true uncertainty of the prediction, with similar
caveats as for a fixed order calculation. The total time required to generate
the event sample of this section, and the next, is roughly one day on a
single PC.

\begin{figure}[tbp]
  \centering
  \includegraphics[width=0.45\textwidth]{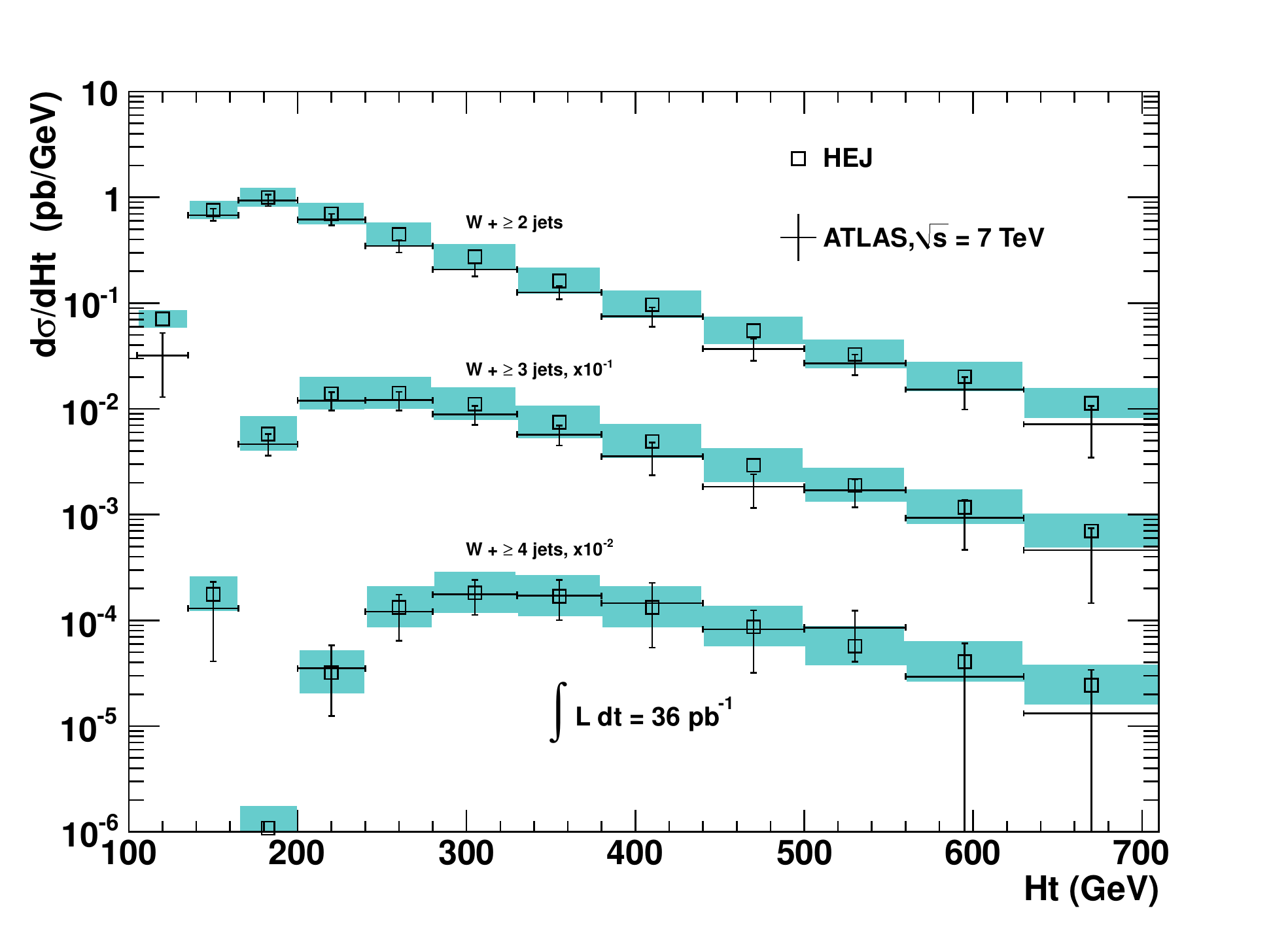}
  \hspace{0.2cm}
  \includegraphics[width=0.45\textwidth]{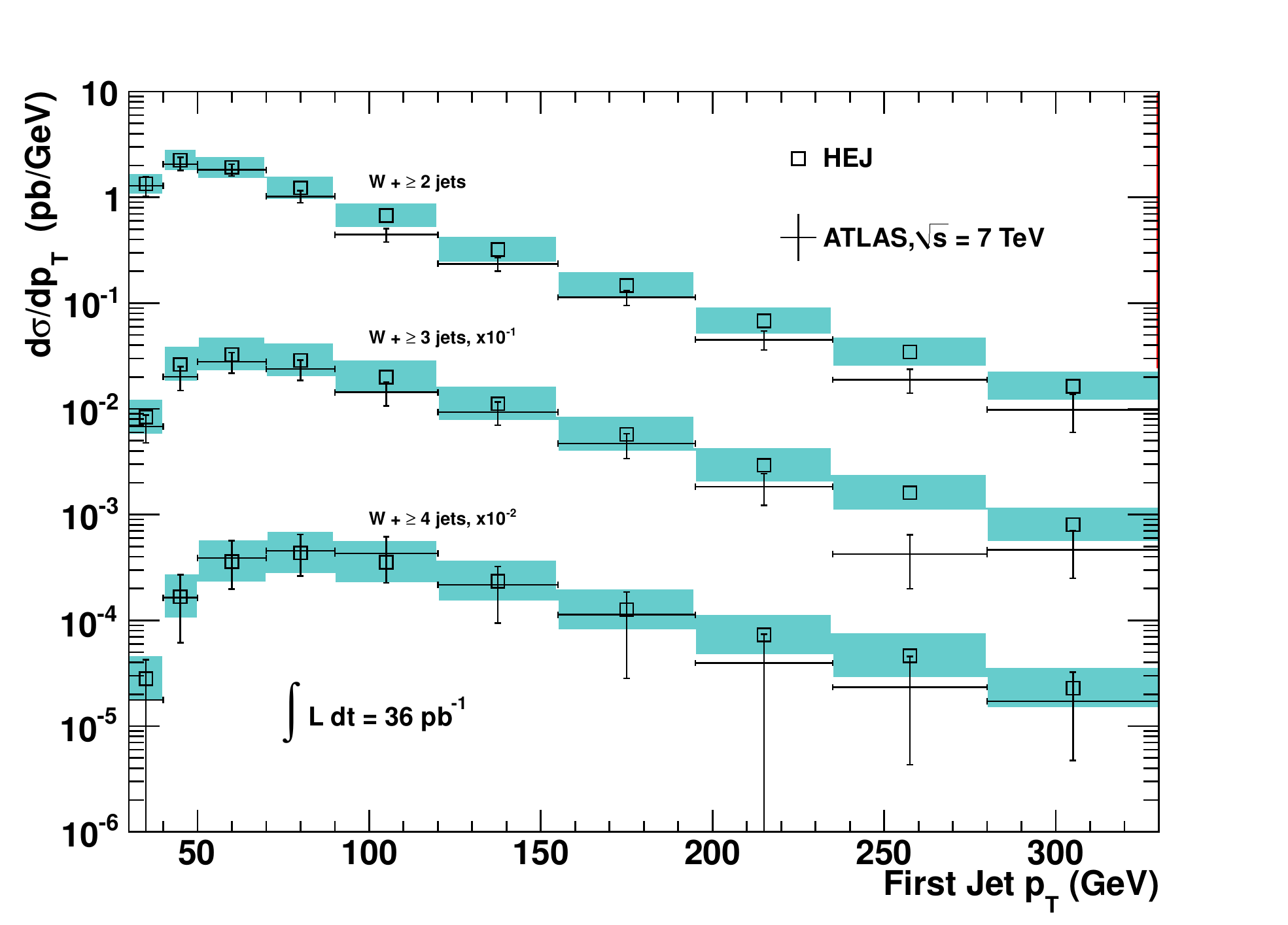}
  \caption{Left: The $W~+$ jets cross section as a function of $H_T$, the scalar sum
    of the transverse momenta of the jets, the charged lepton and the missing transverse
    momentum.  Right: The $W~+$ jets cross section as a function of the transverse
    momentum of the hardest jet in the event.  The data points, in this and
    subsequent plots in this section, are taken from Ref.~\cite{Aad:2012en}.}
  \label{fig:atlasht}
\end{figure}

We begin in figure~\ref{fig:atlasht} by showing comparisons for the
distributions of $H_T$ and the transverse momentum of the hardest jet, in
events of 2,3 and 4 jets.  The left plot shows the $H_T$ distribution for
inclusive $W+2,3$ and $4$-jet samples. In all but the bin of lowest $H_T$ for
the inclusive $Wjj$ production (where the experimental uncertainty is
relatively large), the predictions from \HEJ overlaps with the data, within
the quoted uncertainties. It is clear from this plot that at for $H_T$ larger
than roughly 400~GeV, the suppression from requiring one additional hard jet
is only roughly a factor 2, and not $\alpha_S$. \HEJ is developed
particularly to deal with the case of a large impact from high jet
multiplicity. A comparison between \HEJ and other theoretical descriptions
for this distribution has recently appeared in \cite{Maestre:2012vp}.
The right-hand plot of figure~\ref{fig:atlasht} shows the transverse momentum
distribution of the hardest jet in each event, again for inclusive $W+2,3$ and
$4$-jet events.  This, more inclusive, variable is less sensitive to the
description of additional radiation like that included in \HEJ, and still one
sees good agreement between the \HEJ prediction and data.

\begin{figure}[tbp]
  \centering
  \includegraphics[width=0.45\textwidth]{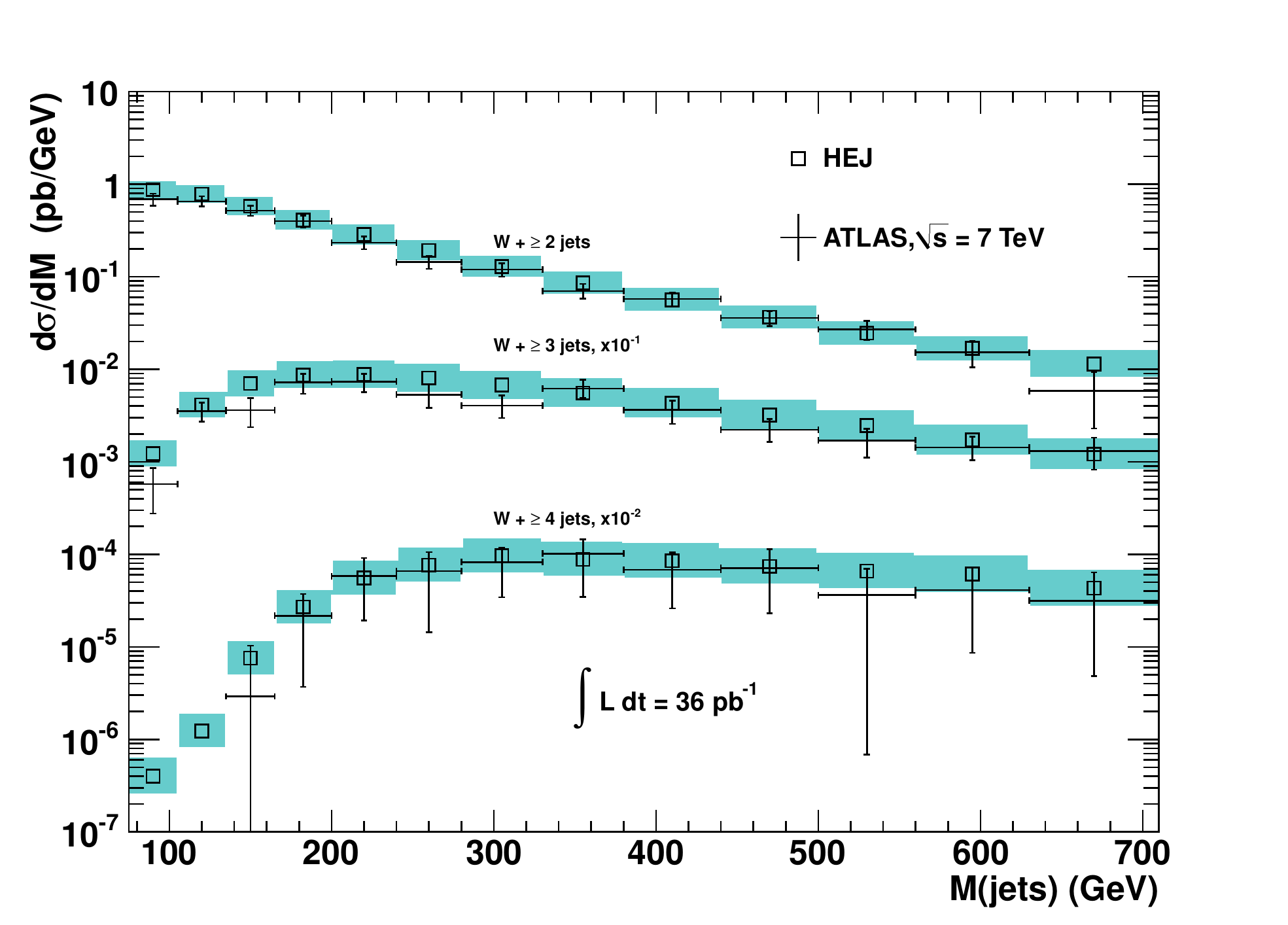}
  \caption{$W~+$ jets cross section as a function of the invariant mass of two, three and four jets separately.}
  \label{fig:mjets}
\end{figure}

Figure~\ref{fig:mjets} shows the distributions for the invariant mass of the
hardest 2, 3 and 4-jets in the event.  This is more sensitive to the topology of
the events, in addition to the overall momentum scale.  As the number of jets
increases, the peak of this distribution moves away from the kinematic minimum
to higher values of invariant mass.  Over this wider range of momentum, we again
find a very close agreement between the \HEJ predictions and the data.

\begin{figure}[tbp]
  \centering
  \includegraphics[width=0.45\textwidth]{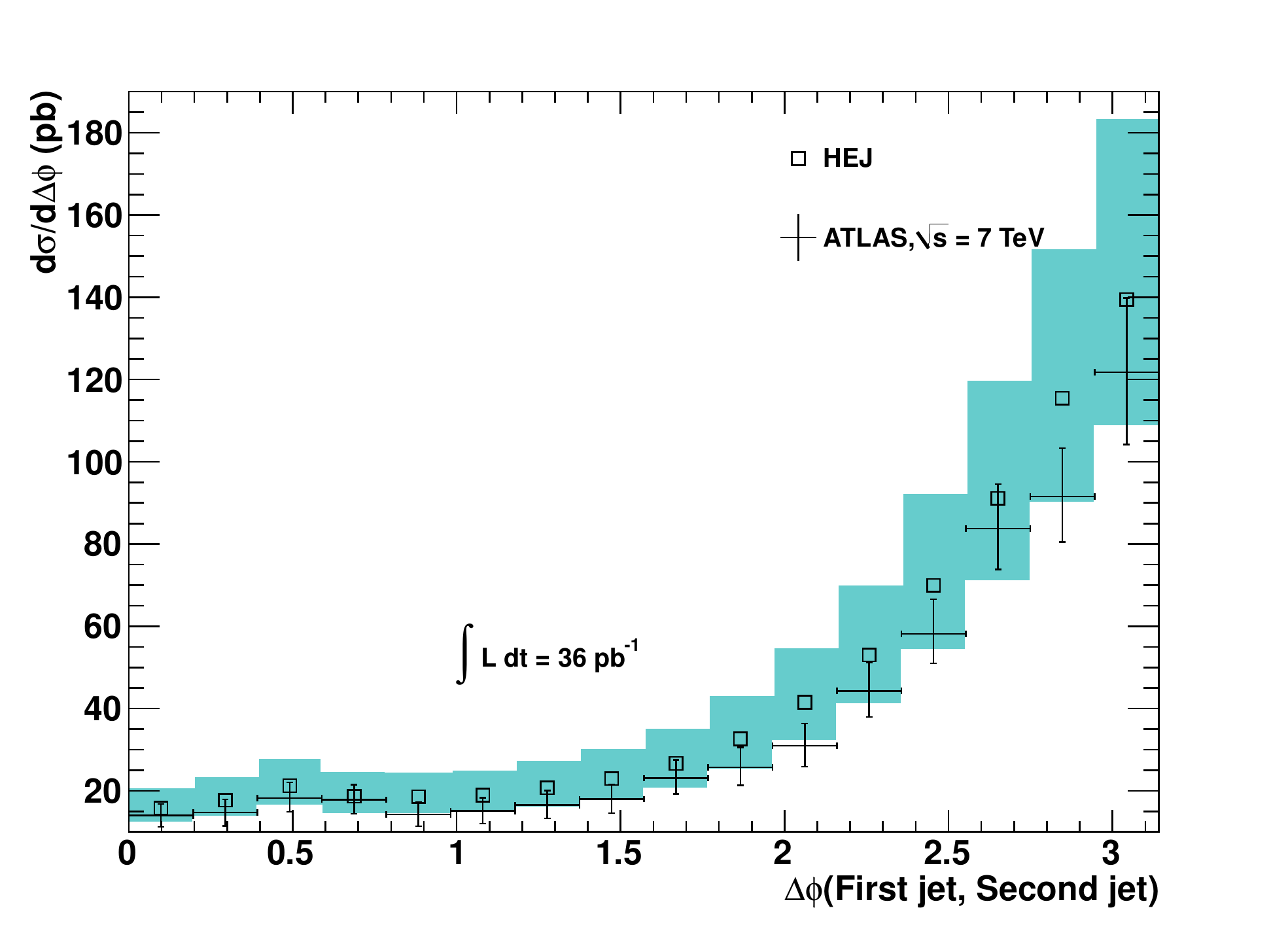}
  \hspace{0.2cm}
  \includegraphics[width=0.45\textwidth]{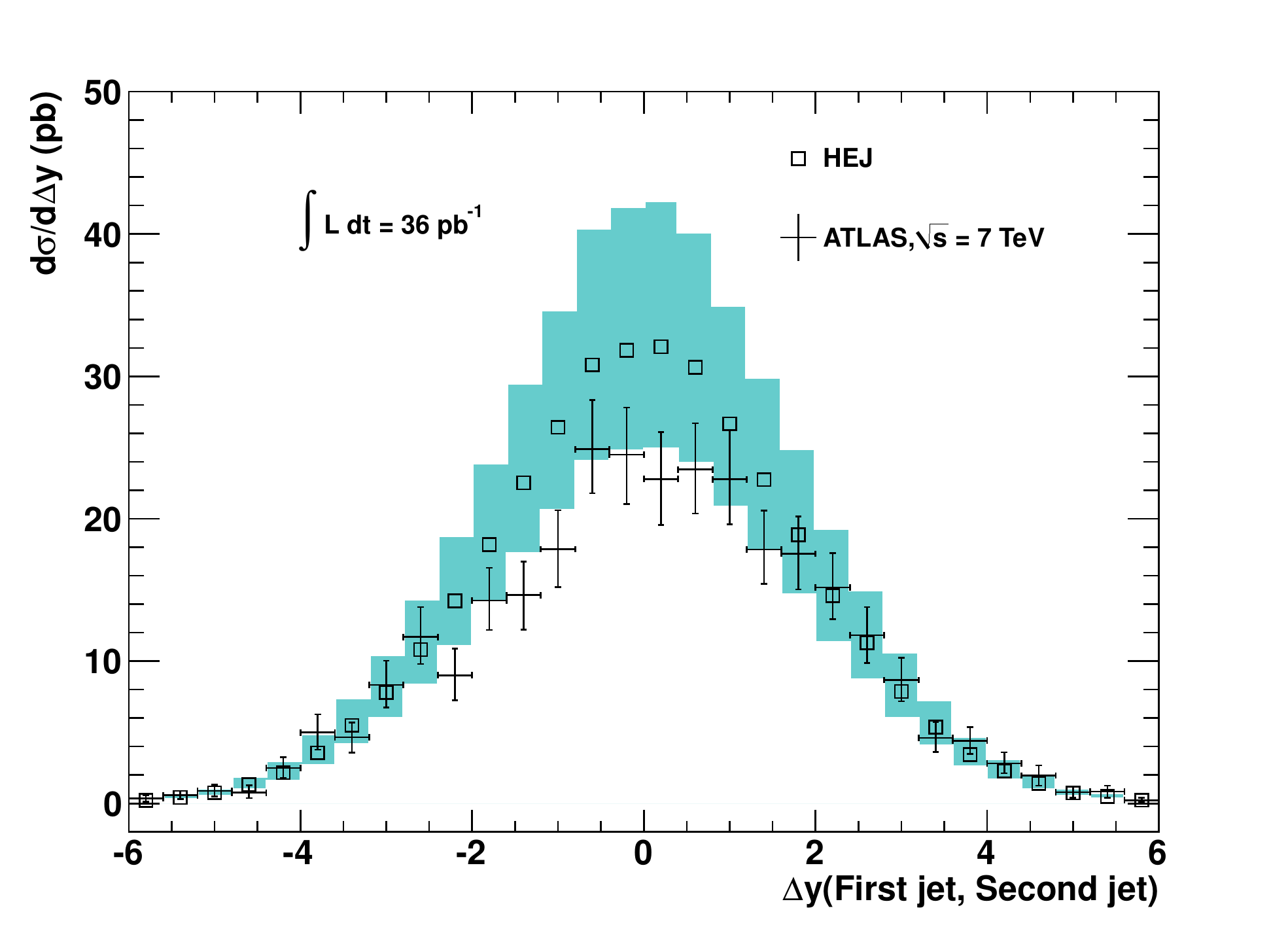}
  \caption{Left: The $W~+$ jets cross section shown as a function of the
    absolute difference in azimuthal angle between the two hardest jets,
    $|\phi_{jet1}-\phi_{jet2}|$.  Right: The $W~+$ jets cross section as a
    function of rapidity difference of the two hardest jets.}
  \label{fig:dydphi}
\end{figure}
The final plots we show in this section, in figure~\ref{fig:dydphi}, probe the
relative position of the two hardest jets in each event.  Firstly, the
left-hand plot shows the distribution of the difference in azimuthal angle between the
two hardest jets.  This is peaked at $\Delta\phi=\pi$. For pure dijet
production at tree-level, the azimuthal distribution is a delta-functional at
$\pi$, and higher order corrections smear out this distribution, which
however remains peaked at $\Delta\phi=\pi$. The fact that data for $Wjj$
shows a similar structure for the azimuthal distribution indicates that this
process proceeds less like one
parton recoiling against a $W$, and then splitting into further jets and more
like a dijet scattering with a $W$-emission. This indeed is the mechanism
implemented in \HEJ, and it describes
the data well across the distribution.  

The right-hand plot in figure~\ref{fig:dydphi} shows the distribution of the
difference in rapidity between the two hardest jets.  It is clear that the
\HEJ description is slightly high at the peak around $\Delta y=0$.  This is
precisely the region where the impact of matching to fixed-order is largest,
and the impact of the na\"ive scheme currently applied will be greatest.
This is an important area of future development for \HEJ.  However, we see
that across the analyses, the description of $W+$jets as currently
implemented in \HEJ describes existing LHC data very well.


\section{Probing Higher Order Corrections}
\label{sec:results}

We have seen that the \HEJ approach gives a good description of the current LHC
data for the production of a $W$ in association with jets.  In this section, we
now turn our attention to variables which may better distinguish between the
predictions obtained in standard approaches (like NLO, a parton shower, or a
combination thereof), and that implemented in \HEJ.

Throughout this section, the \HEJ prediction is shown as the black central
line.  The result of varying the renormalisation and factorisation scale by a
factor of two in either direction is shown by the cyan band. The central
scale choice chosen in this section is $p_{\perp\mathrm{max}}$, the maximum
transverse momentum of any of the jets.

Since all of the remaining studies are of ratios of cross sections with a
correlation between numerator and denominator, the statistical uncertainty
cannot be estimated by the usual propagation of errors derived by assuming no
correlation. Instead, the Monte Carlo uncertainty was evaluated by splitting
our generated events into 12 samples of equal size, and calculating the
distributions from each of the 220 possible ways of selecting 9 out of these
12 samples.  The statistical error band is then defined such that 68\% of
these predictions lie within the central line obtained from all 12 samples.

\begin{figure}[tbp]
  \centering
  \includegraphics[width=0.45\textwidth]{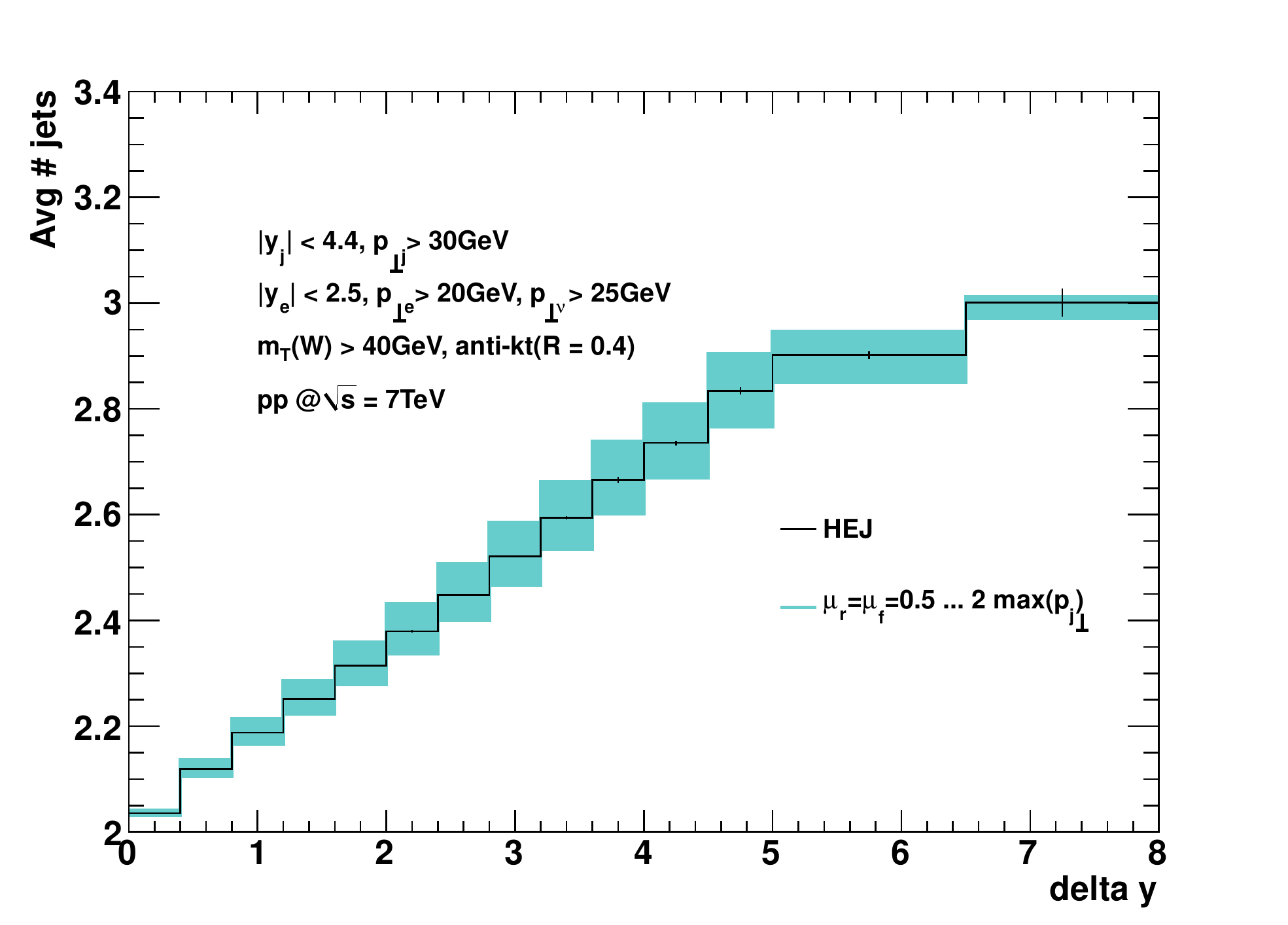}
  \hspace{0.2cm}
  \includegraphics[width=0.45\textwidth]{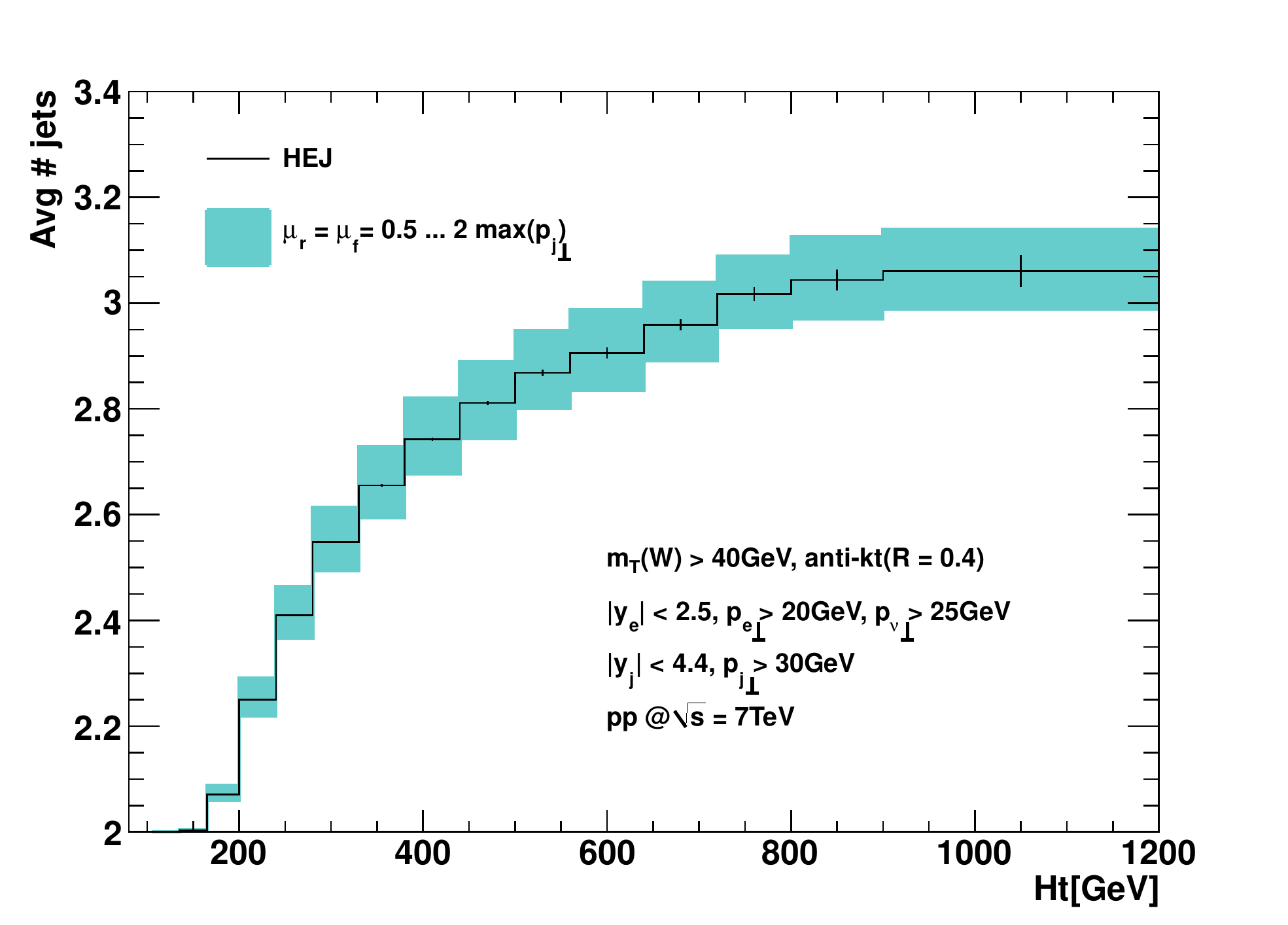}\\
  \includegraphics[width=0.45\textwidth]{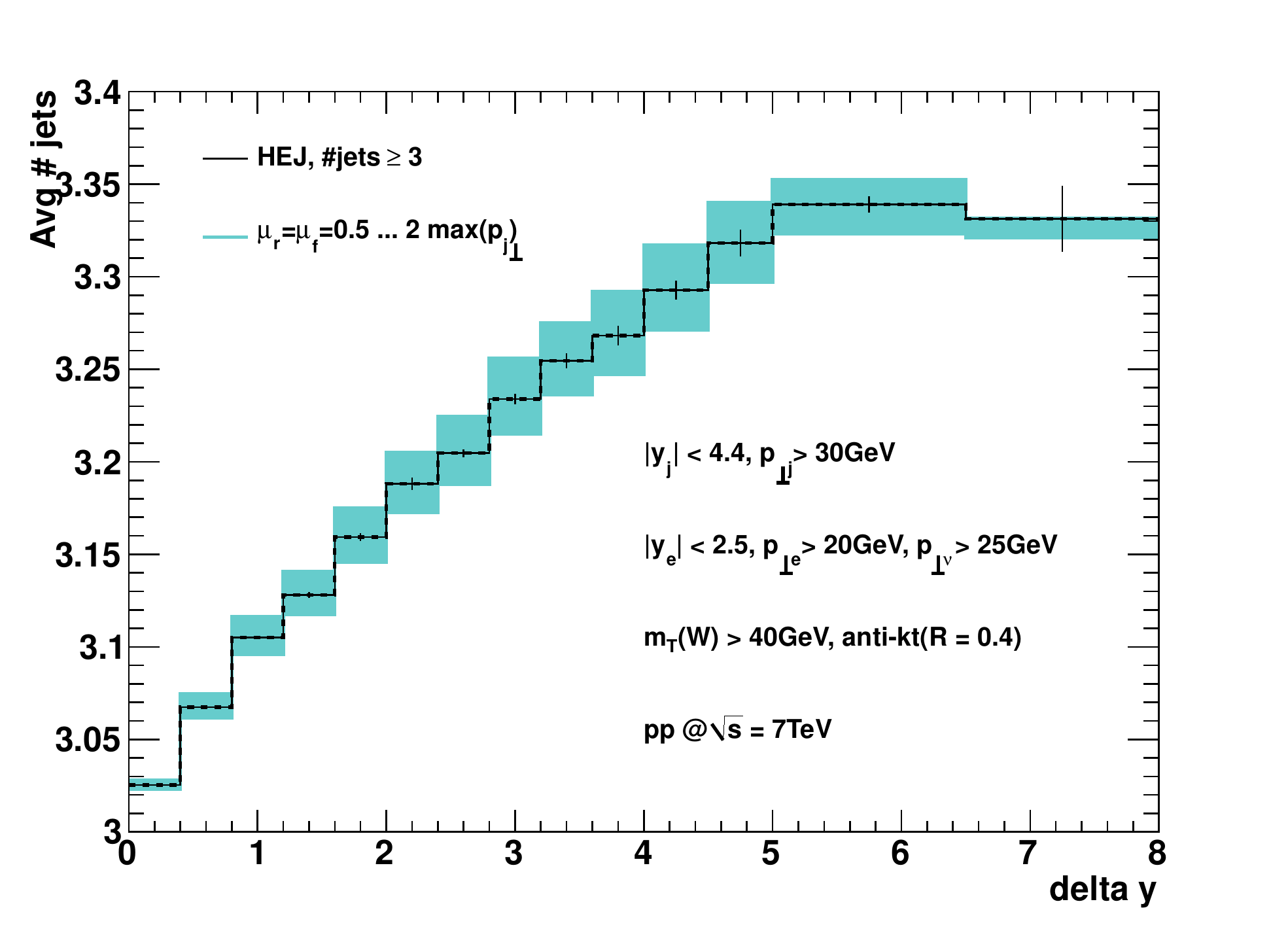}
  \hspace{0.2cm}
  \includegraphics[width=0.45\textwidth]{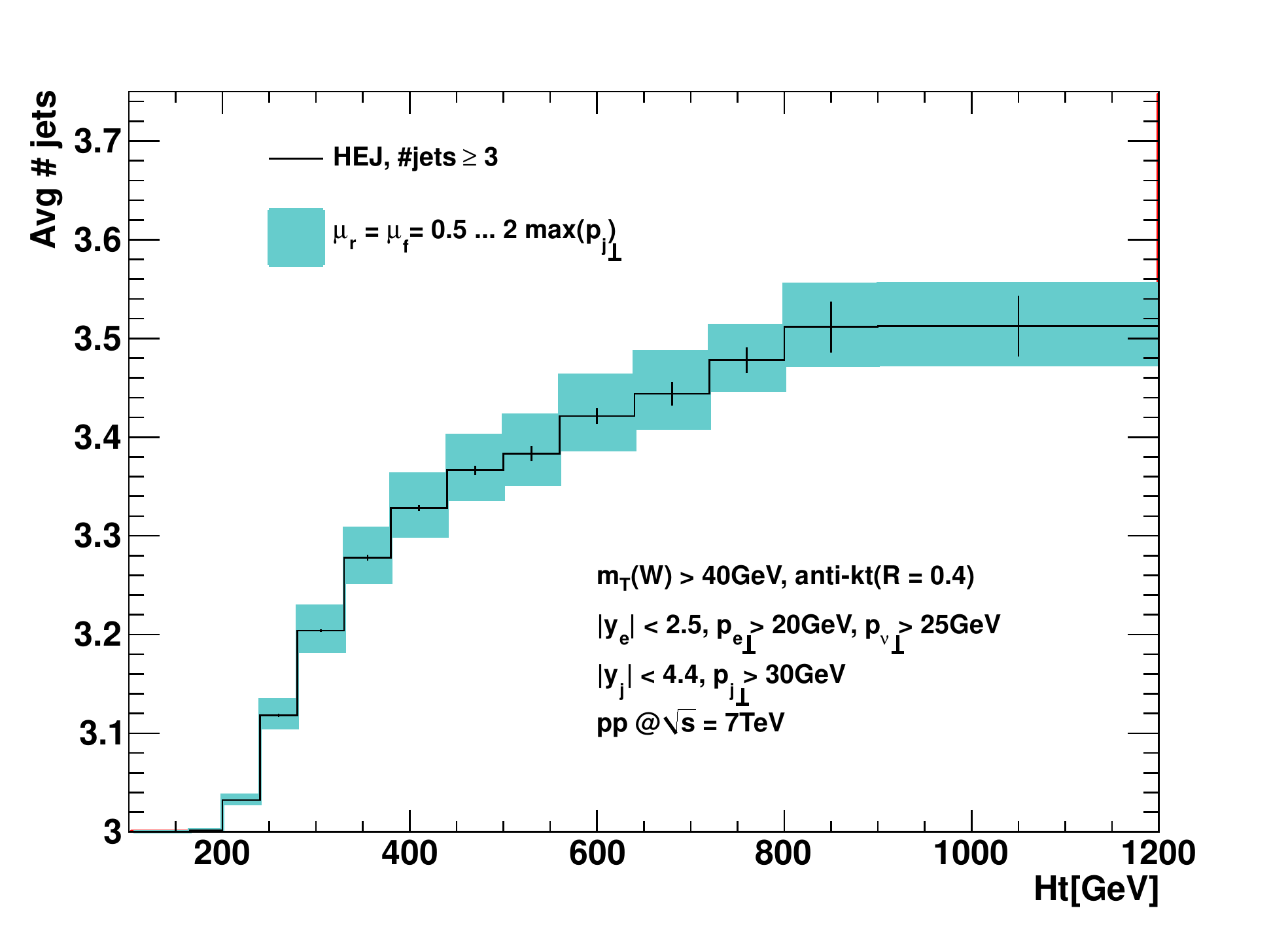}\\
  \caption{Top Left: The average number of jets as a function of the rapidity
    difference between the most forward and most backward jet and Top Right:
    the average number of jets as a function of $H_T$. Bottom Row: as top
    row, except restricted to events with 3 jets or more. }
  \label{fig:njets}
\end{figure}

The first observable we will consider is the average number of observed jets.
The \HEJ predictions for this are shown in figure~\ref{fig:njets} as a
function of $\Delta y$ (left) and $H_T$ (right).  The top plots show results
for inclusive $W+$dijet samples, but the predicted \emph{average} number of
jets in the events rises to around 3 in each case, emphasising the importance
of terms at higher orders in $\alpha_s$ for a reliable jet count. This may
play an important r\^ole in discriminating SM vs.~BSM contributions to the
same channel.  Both of the regions studied here are important regions of
phase space.  This variable has been studied by the ATLAS collaboration in
the context of dijet production~\cite{Aad:2011jz}, where significant effects
beyond fixed order were seen.

In the bottom row of figure~\ref{fig:njets}, the same variable is plotted but
now restricted to events with three resolved jets or more.  Again, we see
that higher orders contribute significantly here, with the average number of
hard jets reaching around 3.3 and 3.5, when shown as a function of $\Delta y$
(the rapidity difference between the most forward and most backward hard jet)
and $H_T$ respectively.

In Ref.~\cite{Maestre:2012vp}, predictions for the average number of jets in
$W+$dijet events were compared between four theoretical approaches: \HEJ, a
pure $Wjj$ NLO calculation\cite{Berger:2009ep,Berger:2009zg,Berger:2010zx},
the ``NLO exclusive sums'' approach (a brute force method to combine NLO
calculations of different orders) and the Sherpa\cite{Gleisberg:2003xi,Gleisberg:2008ta}
MEPS\cite{Hoeche:2009rj,Hoeche:2009xc,Carli:2010cg} scheme which combines
tree-level matrix elements of different orders in $\alpha_s$ with a truncated
parton shower.  Large differences were seen in the predictions between these
theoretical descriptions and an experimental study of this variable would
further our understanding of the nature of QCD radiation in the high-energy
environment of the LHC.

\begin{figure}[tbp]
  \centering
  \includegraphics[width=0.45\textwidth]{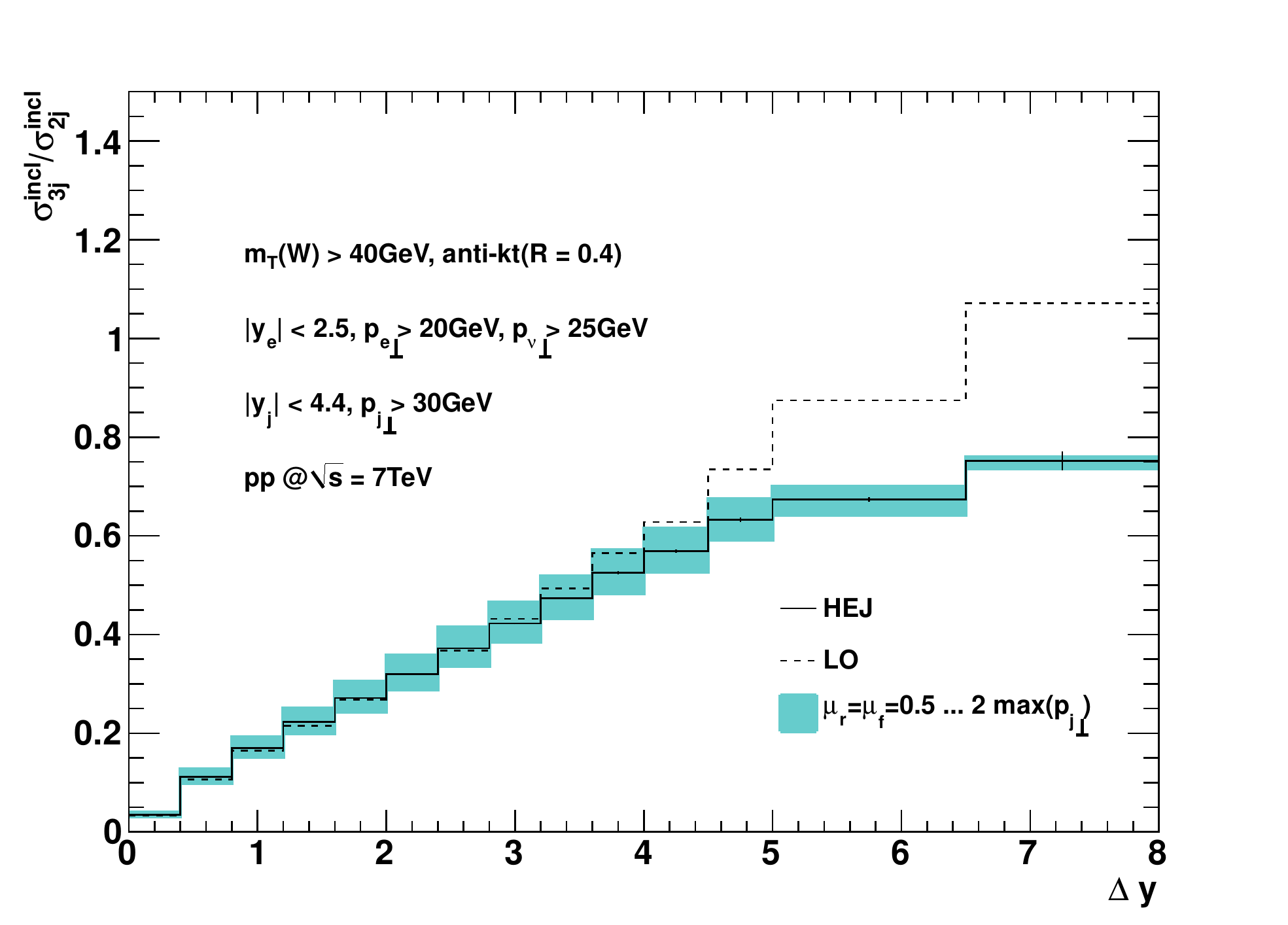}
  \hspace{0.2cm}
  \includegraphics[width=0.45\textwidth]{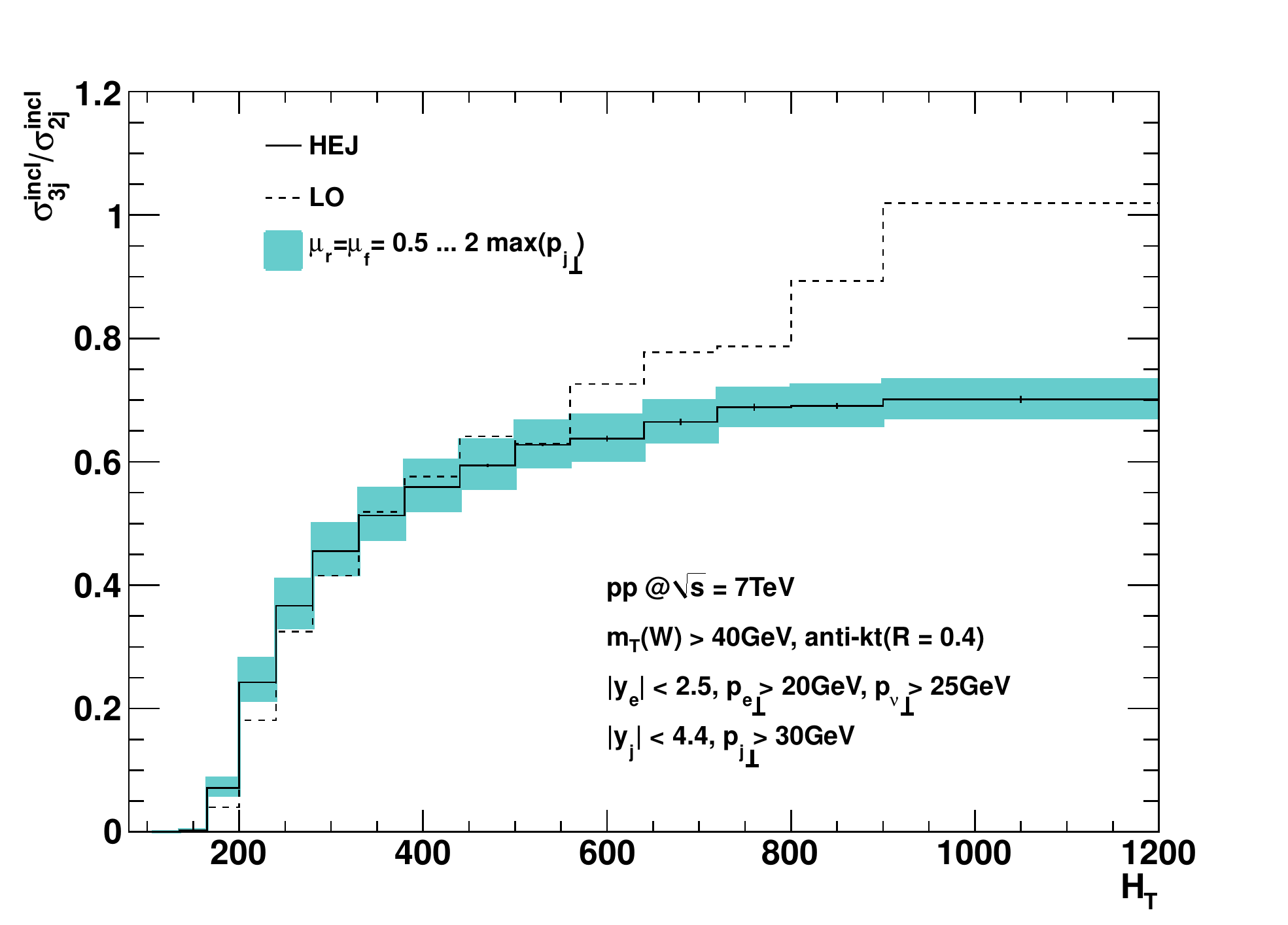}\\
  \includegraphics[width=0.45\textwidth]{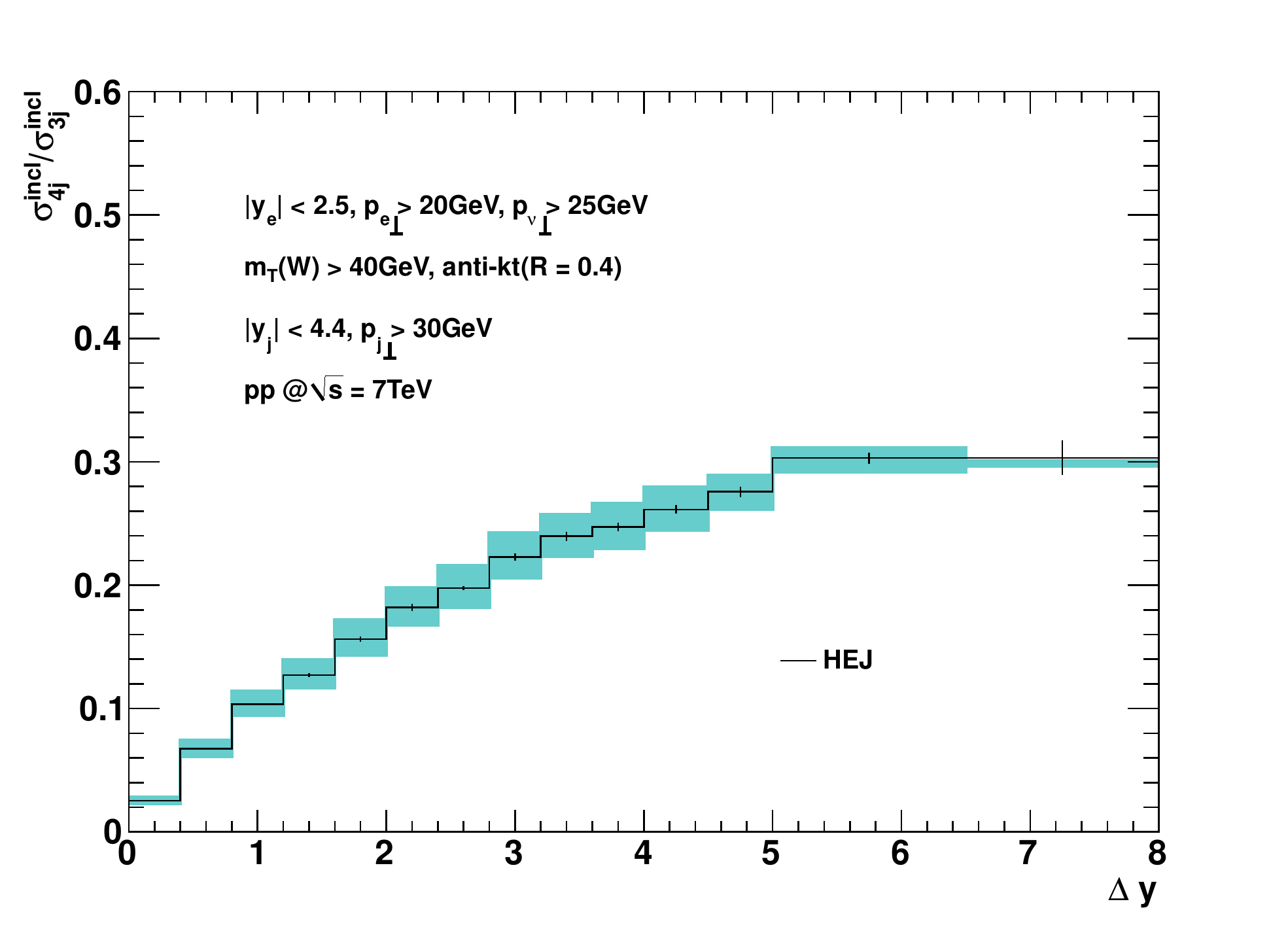}
  \hspace{0.2cm}
  \includegraphics[width=0.45\textwidth]{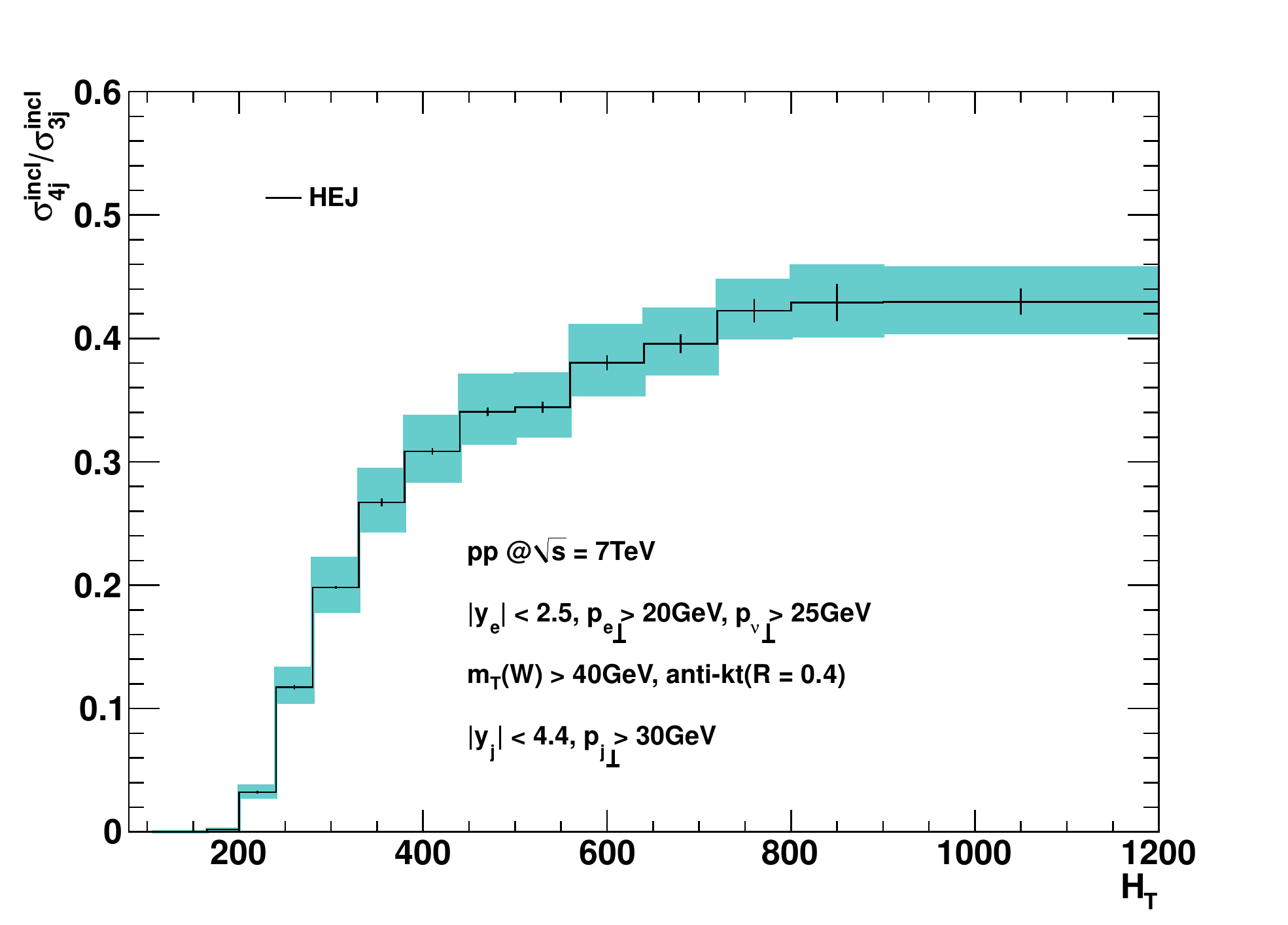}\\
  \caption{Top Left: The ratio between the inclusive 3-jet and 2-jet rates as a
    function of the difference in rapidity between the most forward and most
    backward jet; Top Right: The ratio of the inclusive 3-jet to 2-jet rate as
    function of $H_T$.  The dotted line in each plot shows the ratio between the
    3j and 2j tree-level calculations as a function of $\Delta y$ and $H_T$
    respectively. Bottom Row: as top row, except the ratios are now of the
    inclusive 4-jet rate divided by the inclusive 3-jet rate. }
  \label{fig:incs}
\end{figure}

The ratios of the inclusive jet rates are of course slightly less sensitive
to additional radiation than the average number of jets. In
Fig.~\ref{fig:incs} (top row) we plot the 3-jet to 2-jet rates, along with
the ratio of the inclusive 4-jet to 3-jet rate (bottom row).  Once again it
is clear that the impact from higher orders in large.  For illustration in
the top row, the ratio between the tree-level 3-jet and 2-jet rates has also
been plotted.  Although this contains no systematic resummation of higher
orders, in fact the leading-order result rises higher than that of \HEJ.
The ratio of the inclusive 3-jet to 2-jet rate was also studied in
Ref.~\cite{Maestre:2012vp} for the different theoretical descriptions listed
above.  As expected, smaller differences were seen between the approaches
than in the predictions of the average number of jets, but even so, the
experimental data could probably select a preferred description of the
inclusive jet rates, especially at large $H_T$.

The ratio of the inclusive 4-jet to 3-jet rate is also shown in the bottom
row of figure~\ref{fig:incs}.  The increase here is somewhat smaller than
that seen for the ratio between the 3-jet and 2-jet-rates, but it still rises to
30\% and 45\% as a function of $\Delta y$ and $H_T$ respectively.

\begin{figure}[tbp]
  \centering
  \includegraphics[width=0.5\textwidth]{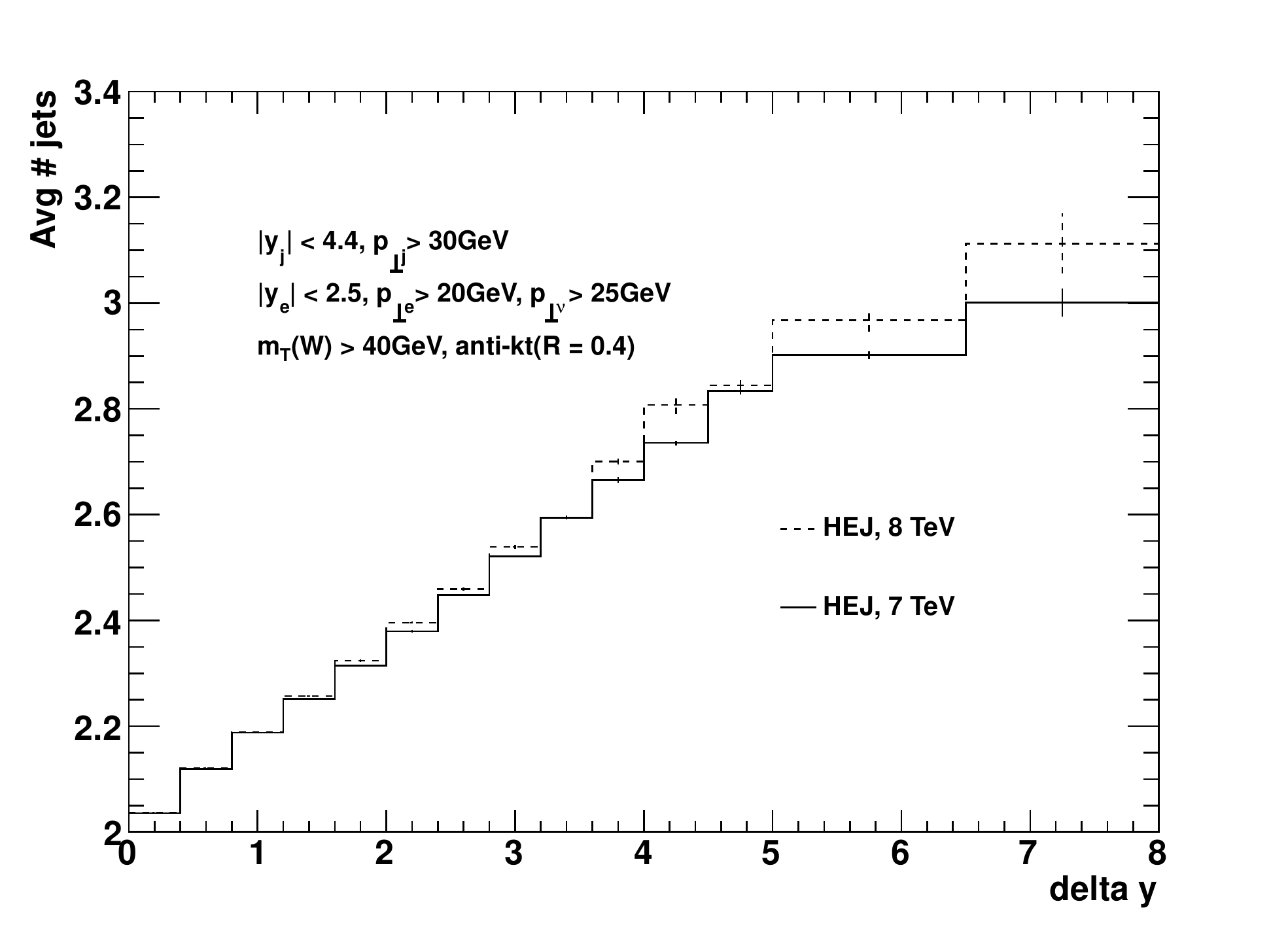}
  \caption{The average number of jets as a function of the rapidity
    span of each event (the difference in rapidity between the most forward and
    most backward jet) for 7 TeV and 8 TeV. }
  \label{fig:7vs8TeV}
\end{figure}

Finally, in Fig.~\ref{fig:7vs8TeV} we compare the predictions from \HEJ for
the average number of jets vs.~$\Delta y$ for inclusive $W+$dijets at the
7~TeV and the 8~TeV LHC. 
The result for 8~TeV shows only a very modest increase in the average number
of jets.

In this section the predictions from \HEJ for the average number of jets and
ratios of inclusive jet rates have been shown.  These show a large degree of
sensitivity to additional hard QCD emissions beyond the leading order.  These
higher-order effects will only increase with the centre-of-mass energy of the
LHC collisions.


\section{Conclusions}
\label{sec:conclusions}

We have described the application of the High Energy Jets (\HEJ) framework to
the production of a $W$ boson in association with at least two jets. \HEJ
resums systematically the contribution from multiple hard emissions
(including also the leading virtual corrections). The process of $W+$dijets
offers a key testing ground for our understanding of the behaviour of the
Standard Model at the LHC, and will in turn be important for many directions
including analyses of Higgs boson couplings and searches for new physics.

The predictions of \HEJ were seen to give a very good description of the
distributions studied with the 2010 data set. We further considered
observables which are designed to be sensitive to the final state
configuration of hard jets, and thus probe the perturbative description.  We
saw that the impact of higher orders on the inclusive dijet sample are large,
for both the average number of jets and the ratios of inclusive jet rates in
various regions of phase space. This offers the possibility of observing
directly the impact of the BFKL-inspired resummation offered by \HEJ.

The implementation of the formalism in the form of a fully flexible partonic
Monte Carlo, can be downloaded at \textsc{http://cern.ch/hej}.

Higher order QCD effects have already been observed in data for pure jet
production, and we look forward to future LHC analyses of $W$+jet production to
further our understanding of physics at these new energy scales.

\section*{Acknowledgements}
\label{sec:acknowledgements}

JRA thanks CERN-TH for kind hospitality while part of this work was performed.
JMS is supported by the UK Science and Technology Facilities Council (STFC).


\bibliographystyle{JHEP}
\bibliography{Wpapers}

\end{document}